\documentstyle[eqsecnum,preprint,aps,epsf]{revtex}
\tightenlines

\newcommand{\nn}{\nonumber}
\def\lsim{\raise0.3ex\hbox{$<$\kern-0.75em\raise-1.1ex\hbox{$\sim$}}}
\def\gsim{\raise0.3ex\hbox{$>$\kern-0.75em\raise-1.1ex\hbox{$\sim$}}}
\newcommand{\ovl}[1]{\overline{#1}}

\newcommand{\eqn}[1]{(\ref{#1})}

\newcommand{\pslash}{p\kern-1ex /}
\newcommand{\Dslash}{{\cal D}\kern-1.5ex /}
\newcommand{\bpsi}{{\overline{\psi}}}
\newcommand{\bq}{{\overline{q}}}

\newcommand{\spr}{{s^\prime}}
\newcommand{\vev}[1]{\left\langle #1 \right\rangle}
\newcommand{\VEV}[3]{\left\langle #1\left| #2 \right| #3\right\rangle}

\begin{document}


\title{
\vspace{-3.0cm}
\begin{flushright}
{\normalsize hep-lat/0007014}\\
{\normalsize UTHEP-428}\\
{\normalsize UTCCP-P-90}\\
\end{flushright}
Chiral properties of domain-wall quarks in quenched QCD}

\author{CP-PACS Collaboration : 
  A.~Ali Khan$^{1}$, 
  S.~Aoki$^{2}$, 
  Y.~Aoki$^{1,2}$, 
  R.~Burkhalter$^{1,2}$, 
  S.~Ejiri$^{1}$, 
  M.~Fukugita$^{3}$, 
  S.~Hashimoto$^{4}$, 
  N.~Ishizuka$^{1,2}$, 
  Y.~Iwasaki$^{1,2}$, 
  T.~Izubuchi$^{5}$, 
  K.~Kanaya$^{1,2}$, 
  T.~Kaneko$^{4}$, 
  Y.~Kuramashi$^{6}$, 
  T.~Manke$^{1}$, 
  K.~I.~Nagai$^{1}$, 
  J.~Noaki$^{1}$,
  M.~Okawa$^{4}$, 
  H.P.~Shanahan$^{1}$, 
  Y.~Taniguchi$^{2}$, 
  A.~Ukawa$^{1,2}$ and 
  T.~Yoshi\'e$^{1,2}$
}
\address{
  ${}^{1}$Center for Computational Physics, University of Tsukuba,
  Tsukuba, Ibaraki 305-8577, Japan\\
  ${}^{2}$Institute of Physics, University of Tsukuba,
  Tsukuba, Ibaraki 305-8571, Japan\\
  ${}^{3}$Institute for Cosmic Ray Research, University of Tokyo,
  Tanashi, Tokyo 188-8502, Japan\\
  ${}^{4}$High Energy Accelerator Research Organization (KEK),
  Tsukuba, Ibaraki 305-0801, Japan\\
  ${}^{5}$Department of Physics, Kanazawa University,
  Kanazawa, Ishikawa 920-1192, Japan\\
  ${}^{6}$Department of Physics, Washington University,
  St.~Louis, Missouri 63130
}

\date{\today}

\maketitle

\begin{abstract}

We investigate the chiral properties of quenched domain-wall QCD (DWQCD)
at the lattice spacings $a^{-1} \simeq 1$  and 2 GeV
for both plaquette and renormalization-group (RG) improved gauge actions.
In the case of the plaquette action we find 
that the quark mass defined through the axial Ward-Takahashi identity
remains non-vanishing in the DWQCD chiral limit that
the bare quark mass $m_f\rightarrow 0$ and the
length of the fifth dimension $N_s\rightarrow\infty$, 
indicating that chiral symmetry is not realized with quenched DWQCD  
up to $a^{-1} \simeq 2$ GeV.
The behavior is much improved for 
the RG-improved gauge action:
while a non-vanishing quark mass remains in the chiral limit 
at $a^{-1}\simeq 1$ GeV, 
the result at $a^{-1}\simeq 2$ GeV is consistent with an 
exponentially vanishing quark mass in the DWQCD chiral limit,
indicating the realization of exact chiral symmetry.
An interpretation and implications are briefly discussed.
\end{abstract}



\section{Introduction}

A recent exciting development in lattice field theory
is the realization of exact chiral-like 
symmetry at finite lattice spacing without fermion doubling.
It started with the domain-wall\cite{Kaplan92,GJK93} and 
the overlap\cite{NN94} formalisms, which were originally proposed 
to formulate chiral gauge theories.   They were soon applied to vector 
gauge theories such as QCD\cite{Shamir93,Shamir95,Neuberger98,HLN98}, 
and progress culminated in the recent discovery\cite{Luscher98} 
that the essential feature of these formalisms is 
the existence of chiral-like 
symmetry which follows from the Ginsparg-Wilson 
relation\cite{Ginsparg}. An explicit form of the fermion action 
satisfying this relation is already known\cite{Neuberger98}.
These developments lead us to expect that QCD with exact chiral 
symmetry can be numerically simulated on a lattice.

The domain-wall fermion formalism is a five dimensional extension of
the Wilson fermion action with a negative mass $-M$.
The theoretical basis is  
that the effective four-dimensional theory obtained by 
integrating out the heavy unphysical modes satisfies the Ginsparg-Wilson 
relation\cite{Neuberger97,KN99}.
On the practical side, the knowledge accumulated on the Wilson
fermion action over the past twenty years enables an efficient
implementation of this system on computers.

There are, however, two subtleties that need to be clarified for
numerical applications of the domain wall formalism to QCD (DWQCD). 
First, the length of the fifth dimension, $N_s$, cannot be
set to infinity:
one has to study the finite $N_s$ effect.
Second, one has to tune the domain-wall height (i.e., fifth dimensional
mass) $M$ to an appropriate value in order to keep the massless mode. 

In the free theory one massless mode exists in the range 
$0<M<2$\cite{Shamir93}.
The two chiral modes, with opposite chiralities, are exponentially bound
to the opposite boundaries in the fifth dimension.
The situation remains unchanged in perturbation 
theory.  An explicit calculation at one-loop order 
shows\cite{AT97,AIKT98} the existence of the massless mode 
guaranteed for the range $0<M'<2$ where $M'$ receives a
shift by the one-loop quantum correction.  It is also easy to
demonstrate that finite $N_s$ effects are suppressed exponentially in
$N_s$.

Numerical tests of chiral properties and physical applications of 
(quenched) DWQCD 
were first discussed in Ref.~\cite{Blum-Soni}. 
Through measurements of pion mass 
and the $K^0-\overline{K}^0$ mixing matrix element  
at the gauge coupling $\beta= 6/g^2\sim6.0$ ($a^{-1}\approx 2$GeV),  
these quantities were shown to vanish in the chiral limit, 
as required by chiral symmetry, 
for a rather small 
fifth dimensional length of $N_s\sim 10$. 
Good chiral behavior has also been observed for $N_s\sim10$ 
for a gauge field background with non-trivial topology\cite{Blum98}.
Even the Atiyah-Singer index theorem has been shown to be approximately
satisfied on 
a lattice for a non-zero quark 
mass\cite{Blum98,LS99}.

These successful results have promoted further applications.
The calculations include
the strange quark mass using nonperturbative renormalization
factors\cite{BSW,RBC9909101,RBC9909107},
the pion decay constant\cite{AIKT9909},
the hadron mass spectrum\cite{RBC9909093,RBC9909117}
and QCD thermodynamics with dynamical
quarks\cite{CV9809159,CV9903024,CV9911002}.

Earlier studies with DWQCD were limited to 
lattice spacings of around $a^{-1}\approx 2$~GeV. 
It is obviously necessary to investigate the  
scaling behavior and carry out the continuum extrapolation. 
Reducing the lattice spacing  
from $a^{-1}\simeq 2$ GeV, however,
is computationally 
difficult, since the formalism requires $N_s$ times more CPU time than 
usual simulations. 
Simulations at coarser lattices are easier, but we must clarify
how large one can increase the lattice spacing while keeping the 
massless mode. 

The chiral properties of quenched DWQCD have been examined in a number of
recent reports 
\cite{RBC9909117,CV9812011,CV9909140,cppacs99}
employing large fifth dimensional lengths up to $N_s\sim50$.
The authors of Refs.~\cite{RBC9909117,CV9812011}, who made a study
with the plaquette gauge action at $\beta=5.7$ and $5.85$ 
($a^{-1}\approx 1-1.5$ GeV), 
claim that pion mass at such strong couplings 
does not vanish in the chiral limit after the $N_s\to\infty$ 
extrapolation, indicating that exact chiral symmetry is not realized.  
Our own study at $\beta=5.65$\cite{cppacs99} yields the same 
conclusion.
We also find that a renormalization-group (RG) improved gluon 
action\cite{Iwasaki83} 
does not improve the situation at $a^{-1}\sim 1$ GeV\cite{cppacs99}.

In this paper we present a more systematic study of 
the chiral property of DWQCD covering both strong and `weak' coupling regions, 
corresponding to lattice spacings $a^{-1}\simeq 1$ and 
2 GeV with both plaquette and RG improved actions.
We emphasize that the chiral
properties are best studied with the quark mass extracted by 
the Ward-Takahashi (WT) identity for the axial vector current.
The difference of this mass and the bare quark mass defines what
we call `anomalous quark mass'.  We discuss that
this quantity suffers from much less systematic uncertainties
than the pion mass from both theoretical and numerical points of view.
We study the behaviors of the anomalous quark mass as a function of
$N_s$ and examine whether it vanishes exponentially in the 
$N_s\rightarrow\infty$ limit for various coupling regimes,
as required by the self-consistency of the proper chiral theory.
This study is also made with varying $M$ and a four dimensional
spatial extent at some fixed coupling. 
We carry out this study with a parallel set of 
simulations employing both plaquette and RG-improved gauge actions 
to see the effect of improvement of the gauge field dynamics 
on the chiral property.
In connection to our work we refer to Ref.~\cite{CV9909140}, which
attempts to extract the WT identity quark mass from the pseudoscalar 
susceptibility.

This paper is organized as follows.
In section~\ref{sec:action} we define the action and the anomalous 
quark mass.  Numerical simulations  and run parameters 
are described in section~\ref{sec:strategy}. 
In section~\ref{sec:plaquette} 
we investigate the chiral property for the plaquette action, and
in section~\ref{sec:RG} for the RG-improved gauge action.
Our provisional interpretation is given in section~\ref{sec:discussion}
for the results we found from the simulation. 
We close the paper with a brief summary and comments on 
the application of DWQCD (section~\ref{sec:conclusion}). 
Appendix A presents an analysis with pion mass, and discuss
whether they support the conclusions based on the anomalous quark mass. 
Numerical data for the anomalous quark mass and pion mass 
are collected in Appendix B.

\section{Action and axial Ward-Takahashi identity}
\label{sec:action}

We adopt Shamir's domain-wall fermion
action\cite{Shamir93,Shamir95},
except that the Wilson term and the domain wall height $M$ have minus signs:
\begin{eqnarray}
S_f &=& -\sum_{x,s,y,\spr} \bpsi(x,s) D_{dwf}(x,s;y,\spr) \psi(y,\spr)
 +\sum_x m_f \bq(x)  q(x)~,
\label{eq:action}
\\
D_{dwf}(x,s;y,\spr) &=& 
 D^4(x,y)\delta_{s,\spr} 
+D^5(s,\spr)\delta_{x,y}
+(M-5)\delta_{x,y} \delta_{s,\spr}~,
\label{eq:qm_dwf}
\\
D^4(x,y) &=& \sum_\mu
\frac{1}{2}\left[(1-\gamma_\mu)U_{x,\mu}\delta_{x+\hat\mu,y}
          + (1+\gamma_\mu)U^\dagger_{y,\mu}\delta_{x-\hat\mu,y}\right],
\\
D^5(s,\spr) &=& \left\{
\begin{array}{lll}
P_L\delta_{2,\spr} &\ & (s=1)\\
P_L\delta_{s+1,\spr} & +P_R\delta_{s-1,\spr} & (1<s<N_s)\\
&P_R\delta_{N_s-1,\spr} & (s=N_s)
\end{array}
\right.~,
\end{eqnarray}
where $x,y$ are four-dimensional space-time coordinates, and $s,s'$ are 
fifth-dimensional or ``flavor'' index,  bounded as 
$1 \le s, s' \le N_s$ with the free boundary condition at both ends
(we assume $N_s$ to be even);
$P_{R/L}$ is the projection matrix $P_{R/L}=(1\pm\gamma_5)/2$, 
and $m_f$ is physical quark mass.

In DWQCD the zero mode of a domain-wall fermion is
extracted by the ``physical'' quark field defined on the edges of the
fifth dimensional space,
\begin{eqnarray}
q(x) = P_L \psi(x,1) + P_R \psi(x,{N_s}),
\nn \\
\ovl{q}(x) = \bpsi(x,{N_s}) P_L + \bpsi(x,1) P_R.
\label{eq:quark}
\end{eqnarray}
The QCD operators are constructed from these quark fields.

For the gauge part of the action we employ the following form in 4
dimensions:
\begin{equation}
S_{\rm gluon} = \frac{1}{g^2}\left\{
c_0 \sum_{plaquette} {\rm Tr}U_{pl}
+ c_1  \sum_{rectangle} {\rm Tr} U_{rtg}
+ c_2 \sum_{chair} {\rm Tr} U_{chr}
+ c_3 \sum_{parallelogram} {\rm Tr} U_{plg}\right\}, 
\label{eqn:RG}
\end{equation}
where the first term represents the standard plaquette action, and the 
remaining terms are six-link loops formed by a $1\times 2$ rectangle, 
a bent $1\times 2$ rectangle (chair) and a 3-dimensional parallelogram. 
The coefficients $c_0, \cdots, c_3$ satisfy the normalization condition
\begin{equation}
c_0+8c_1+16c_2+8c_3=1. 
\end{equation}
The standard Wilson plaquette action is given by
$c_0=1, c_1=c_2=c_3=0$ .
The RG-improved action of Iwasaki\cite{Iwasaki83} is defined by
setting the parameters to $c_0=3.648, c_1=-0.331, c_2=c_3=0$:
with this choice of parameters the action is expected to 
realize smooth gauge field fluctuations approximating those in the continuum 
limit better than with the unimproved plaquette action. 

The chiral transformation of domain wall fermion 
is defined as\cite{Shamir95},
\begin{eqnarray}
&&
\delta\psi(x,s)=iQ(s)\epsilon^a T^a\psi(x,s),
\\ &&
\delta\bpsi(x,s)=-i\bpsi(x,s) Q(s)\epsilon^a T^a,
\end{eqnarray}
where $Q(s)={\rm sign}(N_s-2s+1)$, $T^a$ is a generator of flavor
symmetry group SU$(N_f)$ with $\epsilon^a$ a transformation parameter.
The corresponding WT identity becomes
\begin{eqnarray}
\sum_\mu \vev{\nabla_\mu A_\mu^a(x){\cal O}}
=
2m_f \vev{P^a(x) {\cal O}}
+2\vev{J_{5q}^a(x) {\cal O}}
+i\vev{\delta^a_x {\cal O}},
\label{eqn:WTid}
\end{eqnarray}
where the axial-vector current $A_\mu^a(x)$ and the pseudoscalar 
density $P^a(x)$ are defined by 
\begin{eqnarray}
A_\mu^a(x) &=& \sum_{s} Q(s)
\frac{1}{2}\Bigl(
 \bpsi(x,s) T^a(1-\gamma_\mu) U_\mu(x) \psi(x+\mu,s)
\nn\\&&
-\bpsi(x+\mu,s) (1+\gamma_\mu) U_\mu^\dagger(x) T^a \psi(x,s)
\Bigr),
\\
P^a(x) &=& \bq(x)T^a \gamma_5 q(x),
\end{eqnarray}
and $J_{5q}^a(x)$ representing explicit breaking of chiral symmetry 
takes the form, 
\begin{eqnarray}
J_{5q}^a(x) &=&
-\bpsi(x,{\frac{N}{2}}) T^a P_L \psi(x,{\frac{N}{2}+1})
+\bpsi(x,{\frac{N}{2}+1}) T^a P_R \psi(x,{\frac{N}{2}}).
\end{eqnarray}

For a smooth gauge field background the anomalous contribution 
$\vev{J_{5q}^a(x){\cal O}^b(y)}$ is bounded by an exponentially 
small value with the argument $-N_s$ 
and vanishes in the limit $N_s\to\infty$\cite{Shamir95}.
In this paper we adopt the operator ${\cal O}=P^b(0,{\bf 0})$,
and measure the chiral symmetry breaking effect by
\begin{equation}
m_{5q}=\lim_{t\to\infty}
\frac{\sum_{\bf x}\left<J_{5q}^a(t,{\bf x})P^b(0,{\bf 0})\right>}
	{\sum_{\bf x}\left<P^a(t,{\bf x})P^b(0,{\bf 0})\right>},
\label{eq:m5q}
\end{equation}
which we call `anomalous quark mass'. 
We expect this quantity to be dominated by short-range 
fluctuations since it is determined by the coupling strength of pion 
to the operators $J_{5q}^a$ and $P^b$.  Therefore, 
finite-size effects are likely to be small.  For the same reason, 
there should be 
no quenched chiral singularities in the relation between $m_{5q}$ and 
the bare quark mass $m_f$.  

The axial WT identity is written in terms of the anomalous
quark mass as,
\begin{eqnarray}
\frac{\sum_{\bf x}\nabla_\mu\vev{A^a_\mu(t,{\bf x})P^b(y)}}
 {2\sum_{\bf x}\vev{P^a(t,{\bf x})P^b(y)}}=m_f+m_{5q}.
\end{eqnarray}
The left-hand side of this equation, proportional to the pion mass
squared, vanishes at $m_f+m_{5q} = 0$.

\section{Run parameters and measurements}
\label{sec:strategy}
We explore the chiral properties of DWQCD at both $a^{-1}\simeq 1$ 
and 2~GeV using the plaquette and the RG-improved actions.    
For the plaquette action, we choose the couplings 
$\beta =$ 5.65 and 6.0, which correspond to
$a^{-1}=1.00$ and 2.00 GeV, respectively, if one 
determines the scale from the string tension $\sigma$\cite{scaleP}
assuming $\sigma =$ (440 MeV)$^2$. 
For the RG-improved action, we take $\beta=2.2$ and $2.6$ for which
$a^{-1}=0.97$ and $1.94$ GeV also from the string tension  
reported in Refs.~\cite{IKKY,okamoto,lat99Kaneko}.

We generate quenched gauge configurations on four-dimensional 
$N_\sigma^3\times N_\tau$ lattices. 
One sweep of the gauge update contains one pseudo-heatbath and four 
overrelaxation steps.  For runs at $a^{-1}\simeq 1$~GeV we adopt 
the lattice size of $12^3\times 24$ and $16^3\times 24$ to study  
finite-size effects.  At  $a^{-1}\simeq 2$~GeV only a $16^3\times 32$
lattice is employed.  After a thermalization of 500 (2000) sweeps
hadron propagators are calculated at every 100th  (200th) sweeps 
for $a^{-1}\simeq 1$~GeV (2~GeV).

The domain-wall quark propagator is calculated 
on $N_\sigma^3\times N_\tau\times N_s$ lattices, 
where the gauge configuration on each fifth dimensional coordinate $s$
is identical and is fixed to the Coulomb gauge.  The size of the fifth 
dimension is varied from $N_s=10$ to 50 at $a^{-1}\simeq 1$~GeV, 
and from $N_s=4$ to $24-30$ at $a^{-1}\simeq 2$~GeV, depending on the
gauge action.

To allow a chiral extrapolation, we employ fermion masses of 
$m_f=0.03, 0.05, 0.1$ (at $a^{-1}\simeq 1$~GeV) and 
$m_f=0.02, 0.04, 0.06$ (at $a^{-1}\simeq 2$~GeV).  The domain-wall 
height dependence is examined at $a^{-1}\simeq 1$~GeV
with the choice of $M=1.3, 1.7, 2.1$, and 2.5.  
At $a^{-1}\simeq 2$~GeV we choose $M=1.8$.
 
The propagators are calculated with the conjugate gradient algorithm 
with an even-odd pre-conditioning.  
We place the source at $s=1$ and $N_s$ in the fifth 
coordinate, so that the propagator from the physical field $\overline{q}$ 
to the domain-wall field $\psi_{x,s}$ for arbitrary $(x,s)$ is obtained. 
Both local and exponentially smeared sources are employed in the spatial 
directions, and meson masses are measured for all possible combinations 
of quark masses while only degenerate combinations are evaluated for 
anomalous quark masses.

Our simulation parameters and the number of configurations are
given in Table~\ref{params}.
We note that gauge configurations are generated independently 
for each choice of $N_s$ and $M$, {\it i.e.,} there are no correlations
between the data generated with a different domain-wall height and a
fifth-dimensional size. 

\begin{figure}[p]
 \begin{center}
  \leavevmode
  \epsfxsize=6.7cm \epsfbox{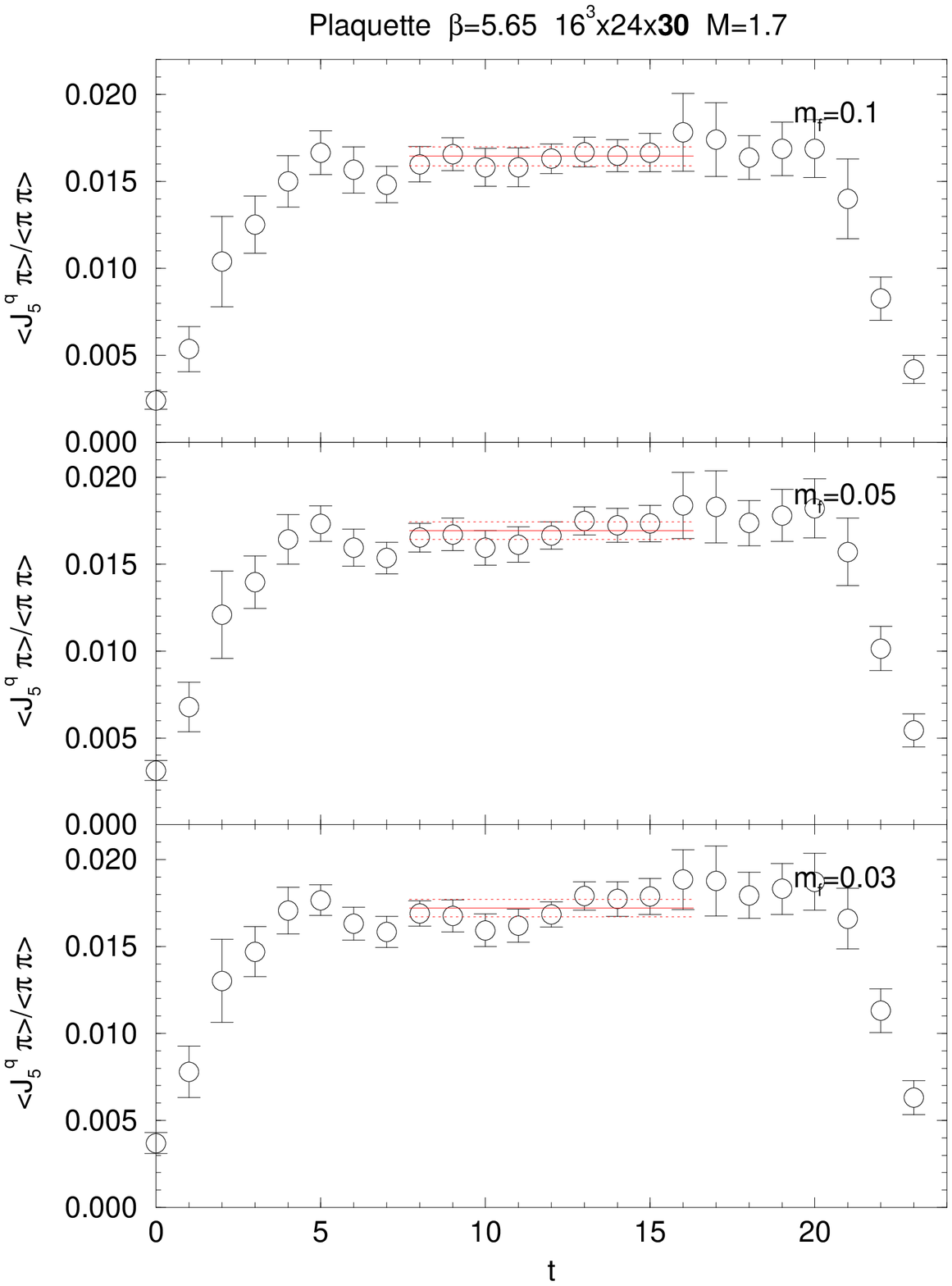}
  \epsfxsize=6.7cm \epsfbox{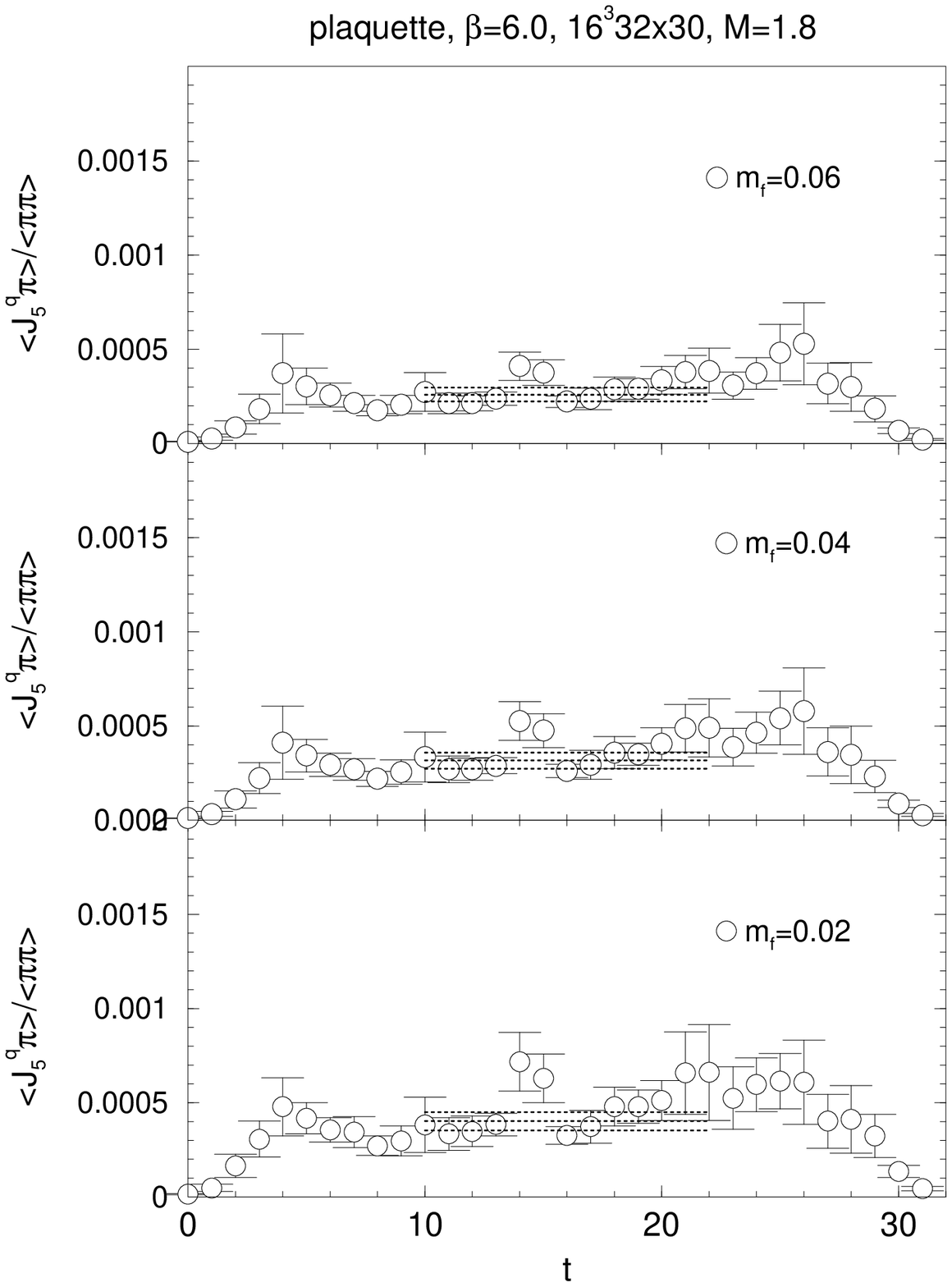}
  \caption{Ratio of two point functions $<J_{5q}(t) P(0)>/<P(t) P(0)>$ 
  for plaquette action at $\beta=5.65$, $M=1.7$ on a $16^3\times24\times30$
  lattice (left) and at $\beta=6.0$, $M=1.8$ on a $16^3\times32\times30$
  (right).  Lines show constant fit over the fitted range.}
  \label{fig:effmqP1}
 \end{center}
\end{figure}

\begin{figure}[p]
 \begin{center}
  \leavevmode
  \epsfxsize=6.7cm \epsfbox{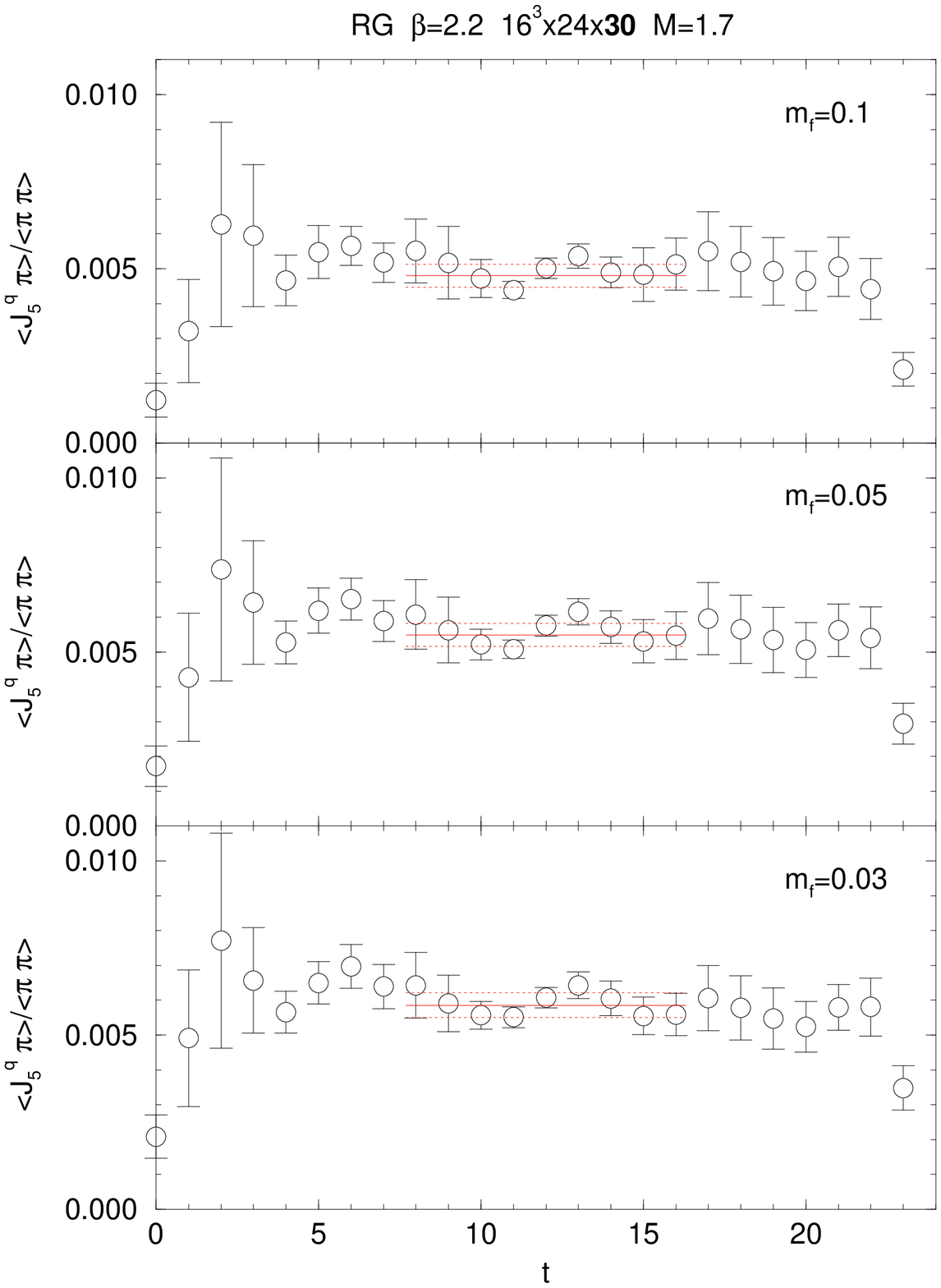}
  \epsfxsize=6.7cm \epsfbox{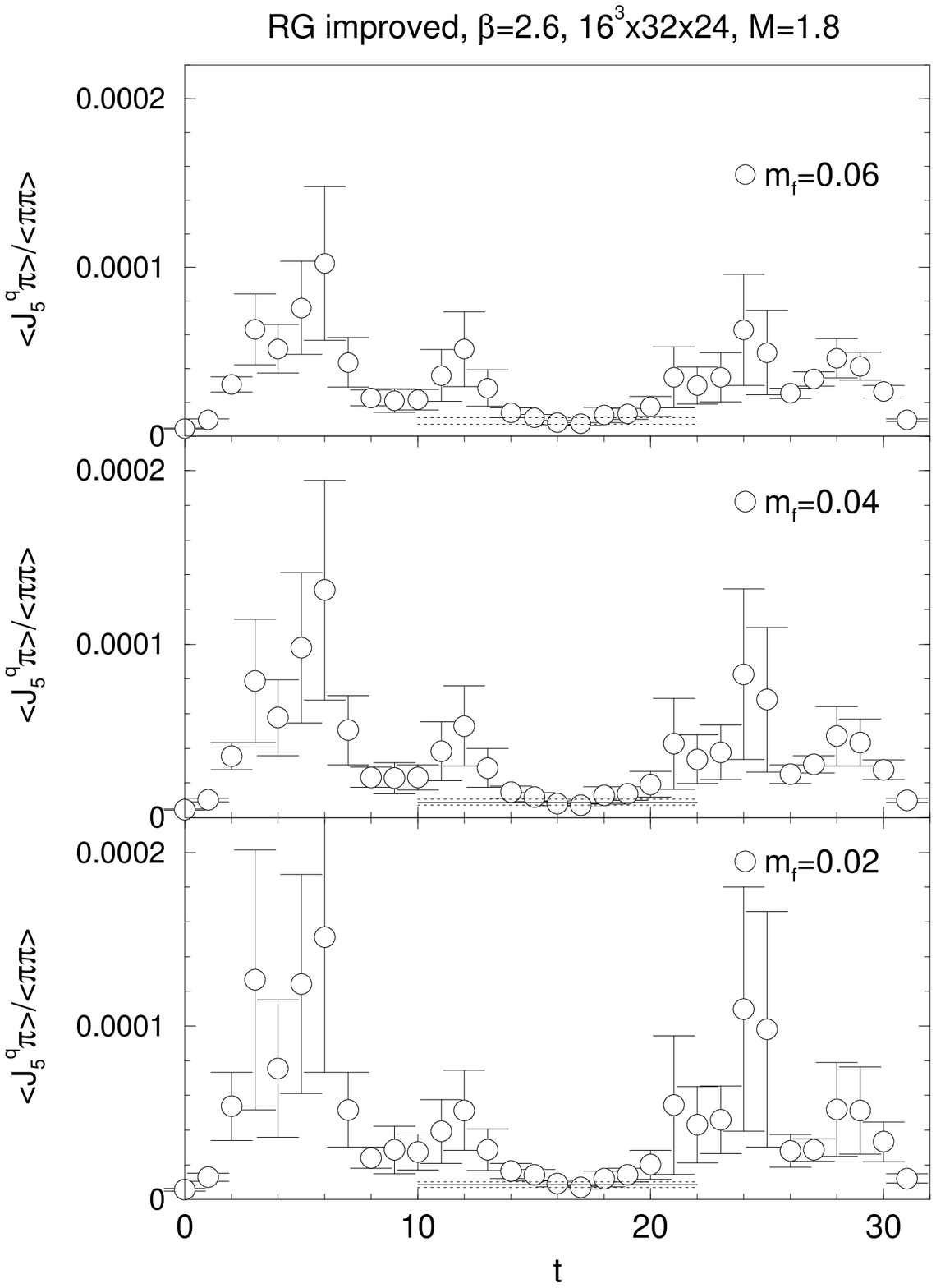}
  \caption{Ratio of two point functions $<J_{5q}(t) P(0)>/<P(t) P(0)>$ 
  for RG improved action at $\beta=2.2$, $M=1.7$ on a
  $16^3\times24\times30$ lattice (left) and at $\beta=2.6$, $M=1.8$ on a 
  $16^3\times32\times24$ lattice (right).
  Lines show constant fit over the fitted range.}
  \label{fig:effmqP2}
 \end{center}
\end{figure}

In Figs.~\ref{fig:effmqP1} and \ref{fig:effmqP2} we  
show typical data for the ratio of two-point 
functions defined in (\ref{eq:m5q}) 
as a function of the temporal distance $t$. 
We obtain the anomalous quark mass $m_{5q}$ at each $m_f, M$ and $N_s$,
by fitting the plateau with a constant, 
the fitting range being determined by the inspection of plots
for the ratio and those for the effective pion mass.
We choose $8\le t \le 16$ as the fitting range for all simulations 
at $\beta=5.65$ (plaquette) and $\beta=2.2$ (RG). We use
$10\le t \le 22$ at $\beta=6.0$ for the plaquette action and 
$\beta=2.6$ for the RG action.
The numerical values for $m_{5q}$ and pion mass are 
given in Appendix B.

\section{Plaquette gauge action}
\label{sec:plaquette}

\begin{figure}[p]
 \begin{center}
  \leavevmode
  \epsfxsize=7cm \epsfbox{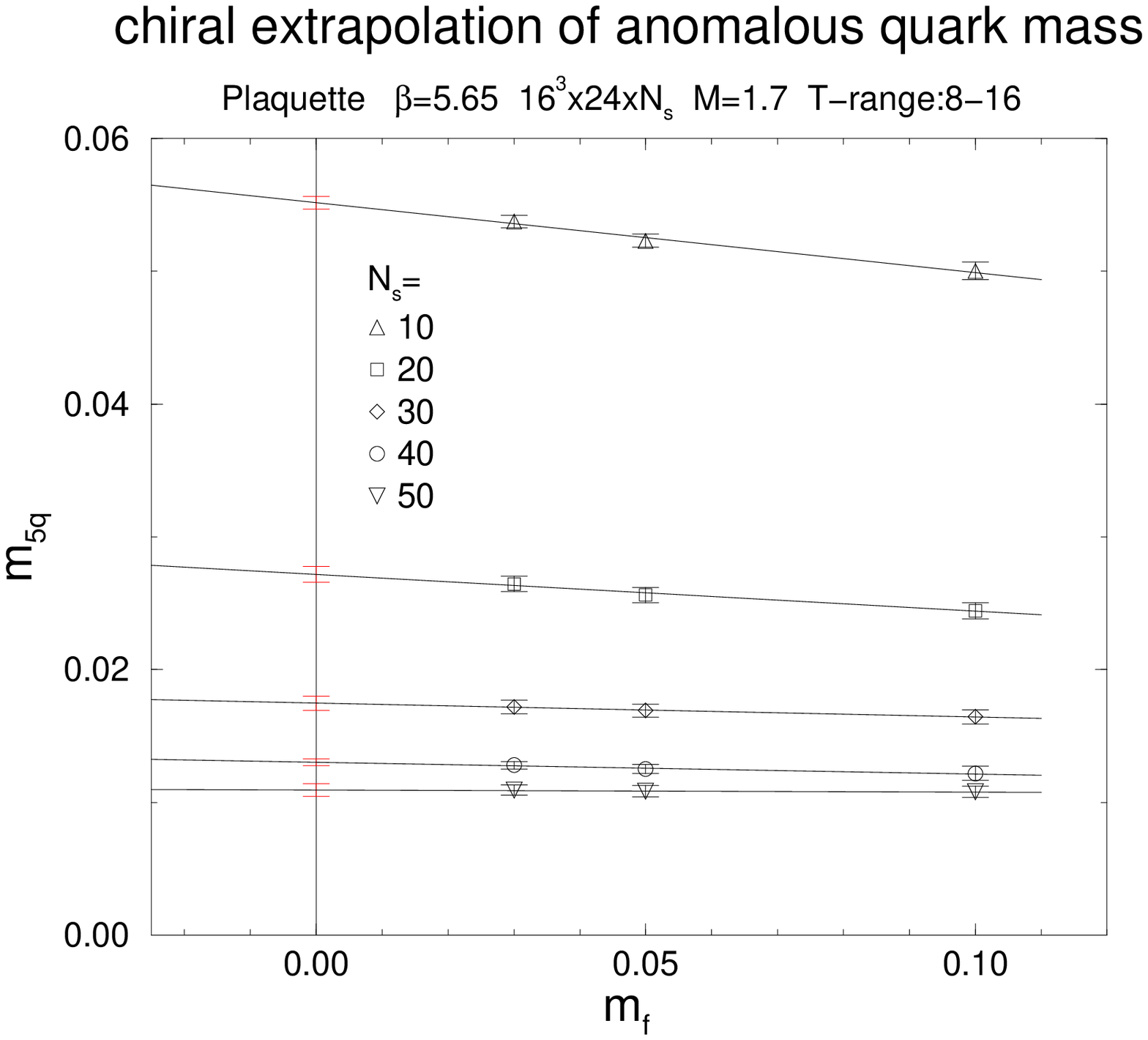}
  \epsfxsize=7cm \epsfbox{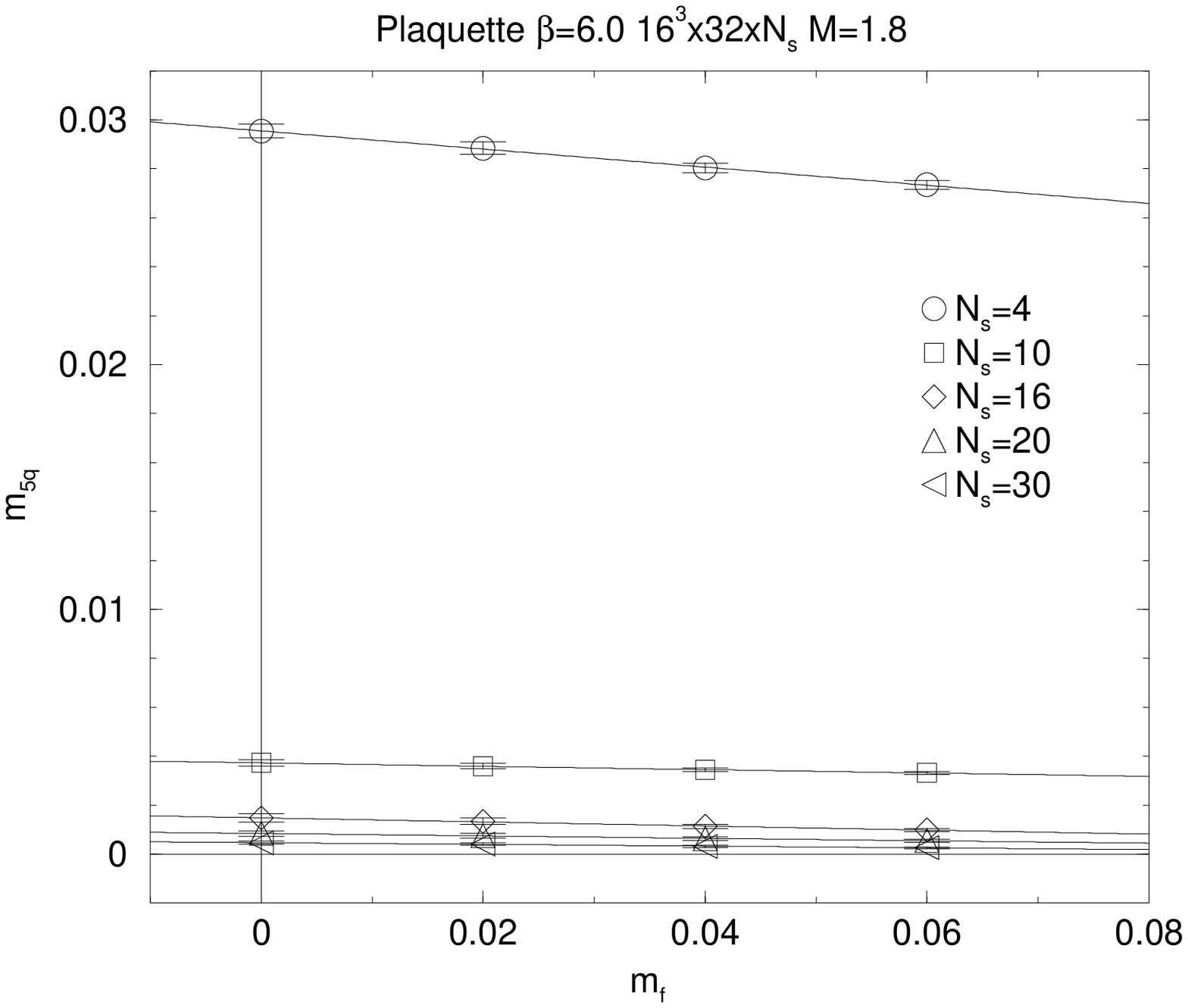}
  \caption{
  Anomalous quark mass $m_{5q}$ as a function of $m_f$ for the
  plaquette action at $M=1.7$ on a $16^3\times24\times N_s$ lattice in the
  strong coupling region (left) and at $M=1.8$ on a $16^3\times32\times N_s$
  lattice in the weak coupling region (right).
}
  \label{fig:m5q-mf-pl}
 \end{center}
\end{figure}

We discuss the anomalous quark mass for the plaquette 
gauge action.  The bare quark mass ($m_f$) dependence of the 
anomalous quark mass $m_{5q}$ is illustrated 
in Fig.~\ref{fig:m5q-mf-pl} for  
$\beta=5.65$ and 6.0. 
We observe only a mild dependence of $m_{5q}$ on $m_f$.   
The anomalous quark mass in the nominal chiral limit
$m_{5q}(m_f=0,N_s,M)$ is obtained by a linear extrapolation in
$m_f$. The extrapolated values are collected in Appendix B.
 
\begin{figure}[p]
 \begin{center}
  \leavevmode
  \epsfxsize=9cm \epsfbox{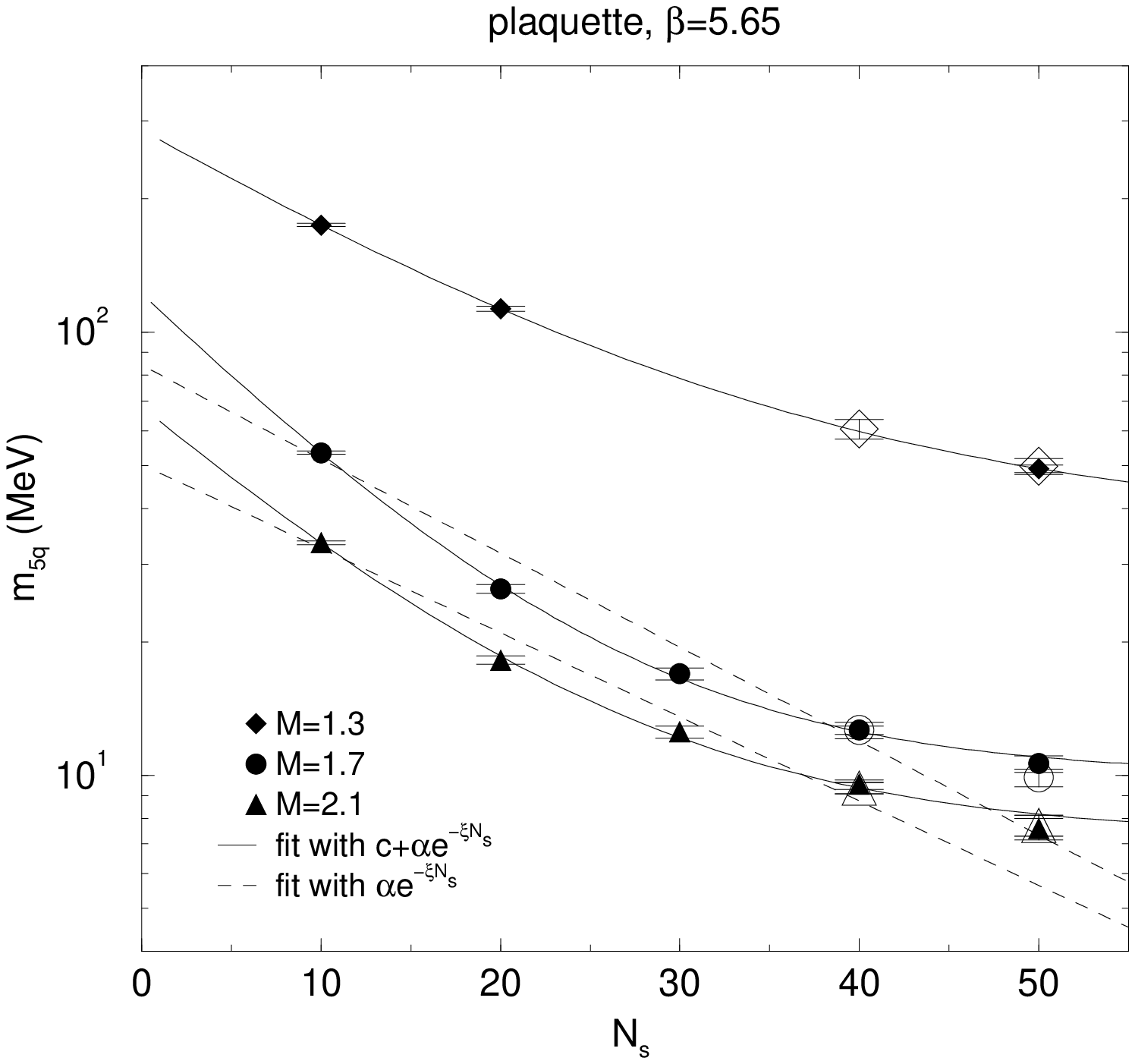}
  \caption{Anomalous quark mass $m_{5q}$ as a function of $N_s$ in
  $m_f\to0$ limit for the plaquette action at $\beta=5.65$.
  Filled symbols are data on $16^3\times24\times N_s$ lattice, and 
  open ones on $12^3\times24\times N_s$ lattice.
  Lines are fits to all the filled points with two functions: 
  $\alpha e^{-\xi N_s}$ (dotted line)
  and $c+\alpha e^{-\xi N_s}$ (solid line).}
  \label{fig:mq0-Ns5-5.65-all}
 \end{center}
\end{figure}

If chiral symmetry is realized in the $N_s \rightarrow\infty$ limit of 
the DWQCD system, $m_{5q}(m_f=0,N_s,M)$ should vanish exponentially in $N_s$. 
We examine this point first at a strong coupling $\beta=5.65$ corresponding to
$a^{-1}\simeq 1$~GeV. 
The anomalous quark mass in the nominal chiral limit, 
$m_{5q}(m_f=0,N_s,M)$, is 
plotted as a function of $N_s$ in Fig.~\ref{fig:mq0-Ns5-5.65-all} 
for three values of $M$ ($M=1.3, 1.7, 2.1$). A comparison of
the filled symbols taken on a  $16^3\times 24\times N_s$ lattice 
with the open ones on 
a spatially smaller $12^3\times 24\times N_s$ lattice shows   
no significant finite-size effect between the two spatial sizes 
corresponding to $N_\sigma a\simeq 2$~ and 3~fm.

The solid lines in Fig.~\ref{fig:mq0-Ns5-5.65-all} show the results of 
fitting data taken on a $16^3\times 24\times N_s$ lattice 
to an exponential with a constant,
$c+\alpha e^{-\xi N_s}$. 
This fit reproduces all five data points with 
an acceptable $\chi^2/{\rm dof}$ for both $M=1.7$ ($\chi^2/{\rm dof}=1.1$) and 
$M=2.1$ ($\chi^2/{\rm dof}=2.2$).  
In contrast an exponential fit without a constant,
$\alpha e^{-\xi N_s}$, 
does not work, resulting in a large $\chi^2/{\rm dof}>30-60$ (dotted lines).

In order to examine the stability of the fit,  
we carry out fits to the four points excluding the data for the smallest 
fifth dimensional length $N_s=10$. 
As seen in Table \ref{tab:m5qfit-pl},
the data are again fitted well with an exponential plus a constant,
but not without a constant.  
The constant $c$ agrees between the five- and four-point
fits within statistical errors.

\begin{figure}[p]
 \begin{center}
  \leavevmode
  \epsfxsize=9cm \epsfbox{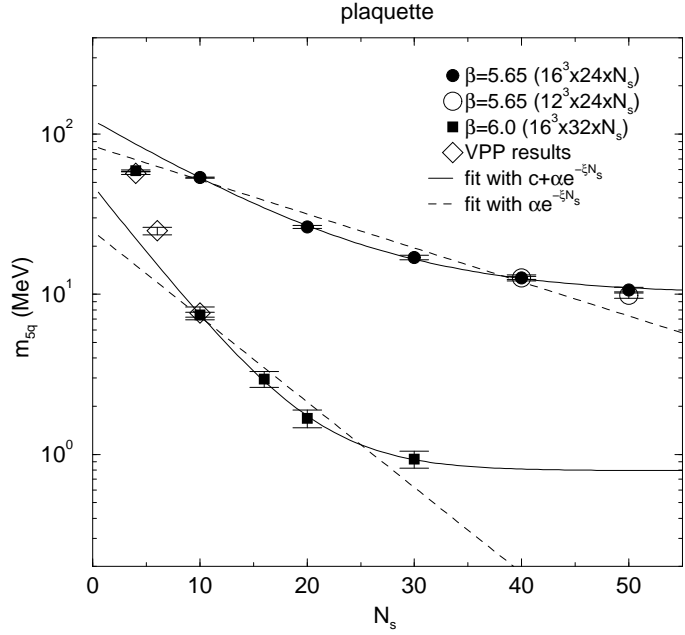}
  \caption{Anomalous quark mass $m_{5q}$ as a function of $N_s$ in
  $m_f\to0$ limit for the plaquette action at weak coupling $\beta=6.0$
  with $M=1.8$ (filled squares) on a $16^3\times 32\times N_s$ lattice. 
  Representative strong coupling results taken at $\beta=5.65$ and $M=1.7$
  are reproduced from Fig.~\protect\ref{fig:mq0-Ns5-5.65-all} for
  comparison (filled and open circles). 
  The results at $\beta=6.0$ from a previous simulation 
  \protect\cite{AIKT9909}
  is also given with open diamonds.
  Lines are fits to filled symbols 
  with two functions: $\alpha e^{-\xi N_s}$ (dotted line)
  and $c+\alpha e^{-\xi N_s}$ (solid line).}
  \label{fig:mq0-Ns5-M17}
 \end{center}
\end{figure}

Our data at $\beta=6.0$ are shown in 
Fig.~\ref{fig:mq0-Ns5-M17} where, for comparison, the
result at $\beta=5.65$ and $M=1.7$ are 
recapitulated from Fig.~\ref{fig:mq0-Ns5-5.65-all}. 
One generally expects that the domain-wall formulation 
works better at weaker couplings\cite{Shamir95}.
Indeed, the decrease of $m_{5q}$ as a function of $N_s$ is much 
more rapid at $\beta=6.0$, giving
$m_{5q}$ an order of magnitude
smaller at $\beta=6.0$ than at $\beta=5.65$:
\begin{eqnarray}
m_{5q}=\cases{
10.32(46)  \;{\rm MeV} & $M=1.7$ at $\beta =5.65$\cr
0.79(16)\;{\rm MeV} & $M=1.8$ at $\beta =6.0$.}
\end{eqnarray}
Nonetheless, we still observe a clear flattening 
of $m_{5q}$ toward a large $N_s$.  
A plain exponential fit without a residual constant does not work.
This is true even if we 
drop the $N_s=4$ data.
The fit including a constant is acceptable
as shown by the solid lines in Fig.~\ref{fig:mq0-Ns5-M17}
(see Table~\ref{tab:m5qfit-pl} for details of $\chi^2$ analyses).

\begin{figure}[p]
 \begin{center}
  \leavevmode
  \epsfxsize=9cm \epsfbox{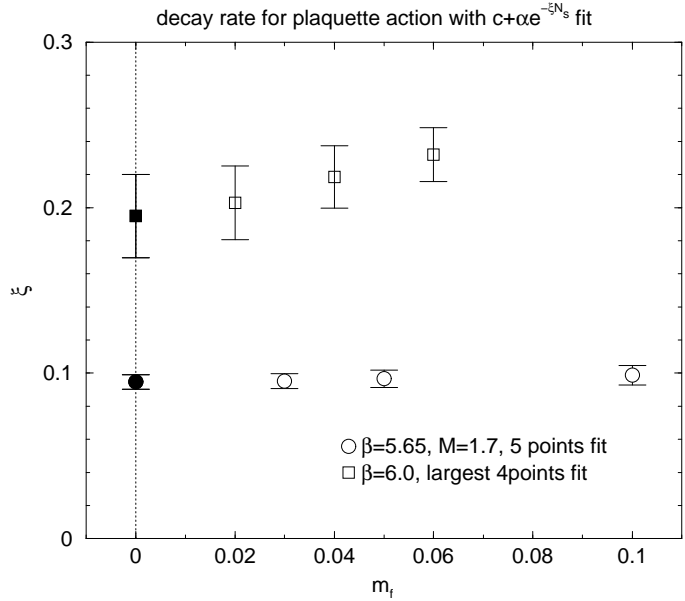}
  \caption{Decay rate $\xi$ of the anomalous quark mass from the
  fit with $c+\alpha e^{-\xi N_s}$ for plaquette action.
  Open circles represent the data in the strong coupling region at
  $M=1.7$, $\beta=5.65$ on $16^3\times24\times N_s$ lattice with five
  points fit.
  Open squares are those in the weak coupling region at $M=1.8$,
  $\beta=6.0$ on $16^3\times32\times N_s$ lattice with four points fit.
}
  \label{fig:drate_mqP5}
 \end{center}
\end{figure}

As a further test for the presence of a non-zero constant,
we extract the decay rate $\xi$ in $N_s$ not only at
the chiral limit ($m_f=0$) but also at a finite $m_f$.
According to the transfer matrix description of 
DWQCD\cite{Shamir95,Neuberger97,KN99},
the mass term, which appears only at the boundary in the fifth dimension,
does not affect the large $N_s$ dependence of $m_{5q}$; 
hence we expect that $\xi$ does not depend on $m_f$.
The decay rates  $\xi$ from a fit with an exponential and a constant
are plotted in Fig.~\ref{fig:drate_mqP5}, which show that 
this expectation is well satisfied at a strong coupling, and within 
estimated errors also at a weaker coupling.  
The decay rate extrapolated to the chiral limit from 
finite $m_f$ is consistent with the value directly 
extracted for $m_f=0$, showing the consistency of the analysis. 

We conclude that a residual constant remains in the anomalous quark mass
in the $N_s\rightarrow\infty$ limit.
The presence of a residual constant, even if it is small,
means that quenched DWQCD does {\it not} realize the expected chiral symmetry 
for the plaquette action, at least, up to 
$a^{-1}\simeq 2$ GeV.

We remark that our conclusion differs from Ref.~\cite{AIKT9909}, which
indicates that $m_{5q}$ vanishes 
exponentially in $N_s$ at $\beta =6.0$ with the plaquette action.
We find that the two data for $m_{5q}$ are mutually consistent 
in the overlapping
range of $N_s$ (see Fig.~\ref{fig:mq0-Ns5-M17}, where open points are
from Ref.~\cite{AIKT9909} and the solid points denote our data). 
The choice of  $N_s\leq 10$ in  Ref.~\cite{AIKT9909}
was too small to observe the asymptotic flattening.

\section{Renormalization group improved action}
\label{sec:RG}

We may ascribe the failure in fulfilling chiral symmetry seen in 
the last section to the roughness of gauge configurations with the plaquette 
action at the lattice spacings we have studied.
We suspect that the massless mode may exist on 
sufficiently smooth gauge configurations close to the continuum 
limit \cite{HJL,Kikukawa}. 
One way to realize smooth gauge configurations, yet keeping the lattice
spacing coarse enough to make a computation feasible,
is to employ improved gauge actions. So 
we discuss the case with the RG-improved 
gauge action\cite{Iwasaki83} in this section. 

\begin{figure}[p]
 \begin{center}
  \leavevmode
  \epsfxsize=9cm \epsfbox{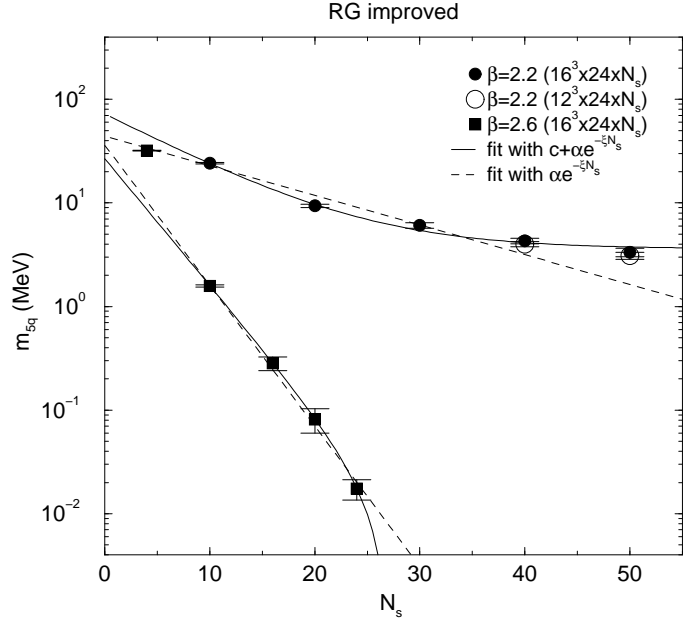}
  \caption{Anomalous quark mass $m_{5q}$ as a function of $N_s$ in
  $m_f\to0$ limit for the RG improved action.
  Filled circles represent data in the strong coupling region at
  $M=1.7$ and $\beta=2.2$ on $16^3\times24\times N_s$ lattice.
  Filled squares are those in the weak coupling region at $M=1.8$ and 
  $\beta=2.6$ on $16^3\times32\times N_s$ lattice.
  Lines are fits with two functions: $\alpha e^{-\xi N_s}$ (dotted line)
  and $c+\alpha e^{-\xi N_s}$ (solid line).
  Data for $N_s=10-50$ (5 points) and $10-24$ (4 points) are used for the fit
  in the strong and weak coupling. }
  \label{fig:mq0-Ns5-M17-R}
 \end{center}
\end{figure}

\begin{figure}[p]
 \begin{center}
  \leavevmode
  \epsfxsize=9cm \epsfbox{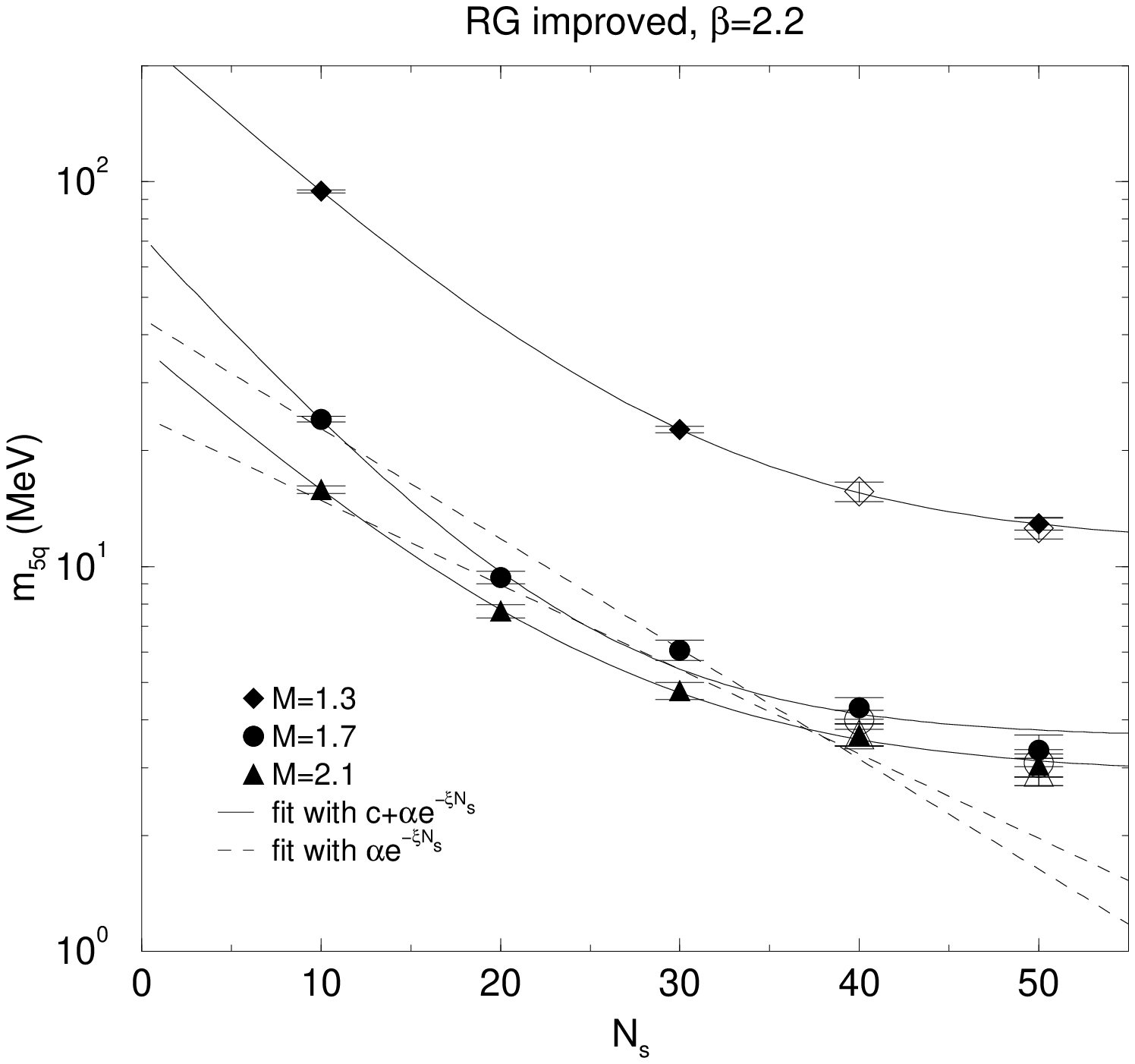}
  \caption{Anomalous quark mass $m_{5q}$ as a function of $N_s$ in
  $m_f\to0$ limit for the RG-improved action at $\beta=2.2$.
  Filled symbols are data on $16^3\times24\times N_s$ lattice, and 
  open ones on $12^3\times24\times N_s$ lattice.
  Lines are fits to all the filled points with two functions: 
  $\alpha e^{-\xi N_s}$ (dotted line)
  and $c+\alpha e^{-\xi N_s}$ (solid line).}
  \label{fig:mq0-Ns5-2.2-all}
 \end{center}
\end{figure}

The anomalous quark mass
in the nominal chiral limit $m_{5q}(m_f=0,N_s,M)$ is
plotted as a function of $N_s$ in Figs.~\ref{fig:mq0-Ns5-M17-R} and 
\ref{fig:mq0-Ns5-2.2-all}. 
Fig.~\ref{fig:mq0-Ns5-M17-R} compares a typical  
result at a strong coupling ($\beta=2.2$) with that 
at a weaker coupling ($\beta=2.6$).
Fig.~\ref{fig:mq0-Ns5-2.2-all} displays the results at $\beta=2.2$
for three different choices of $M$ and for two different choices of 
the spatial lattice size $N_\sigma$.
A comparison of solid points 
($16^3\times 24\times N_s$ lattices)
and open points ($12^3\times 24\times N_s$)
verifies the absence of a significant finite size effect at this $\beta$. 
In order to take the $N_s\to\infty$ limit we carry out 
exponential fits, with or without a constant, as we did for the  
plaquette action. The fit parameters are summarized in Table
\ref{tab:m5qfit-rg}.

We first discuss the results in the weak coupling region displayed 
in Fig.~\ref{fig:mq0-Ns5-M17-R}.
As one can see from large values of $\chi^2/dof$ in  
Table~\ref{tab:m5qfit-rg}, 
neither $c+\alpha e^{-\xi N_s}$ $(\chi^2/{\rm dof}=12.8)$ nor 
$\alpha e^{-\xi N_s}$ $(\chi^2/{\rm dof}=16.9)$
fits the data well in so far as the point at $N_s=4$ is included.
This means that the leading exponential term does not dominate 
the two point correlators
owing to a non-trivial overlap of quark 
wave functions between the two domain walls for this small $N_s$.

If the data at $N_s=4$ are excluded, however, 
the four points for $N_s\geq 10$ are 
fitted equally well with either $c+\alpha e^{-\xi N_s}$ (solid line) or
$\alpha e^{-\xi N_s}$ (dotted line) as seen in the figure.
In fact,
the fit with a constant yields $c=-0.71(63)\times 10^{-5}$, 
consistent with zero at one standard deviation;
the four data points fall on
$\alpha e^{-\xi N_s}$, indicating the realization of chiral symmetry.

We obtain a different result for a strong coupling. Solid lines in  
Fig.~\ref{fig:mq0-Ns5-2.2-all}
show fits to all five data points  
with an exponential plus a constant.
They provide reasonable fits to all data. 
On the other hand, a plain exponential
does not fit the data.

Repeating the analysis for the four data points excluding $N_s = 10$, 
we still find the fit with a non-zero constant ($\chi^2/{\rm dof}=0.1$) 
being much better than a plain exponential ($\chi^2/{\rm dof}=4.0$) at
$M=2.1$.  The constant $c$ is consistent 
between the four- and five-point fits.  This analyses 
supports the conclusion that chiral symmetry is not realized at $M=2.1$. 
The four data points at $M=1.7$, on the other hand, are fitted well with
either of the two forms with $\chi^2/{\rm dof}<1$. 
The constant $c$ for the four-point fit is 
consistent with zero within two standard deviations.
Thus the possibility that $m_{5q}$ vanishes exponentially in $N_s$ 
at $M=1.7$ cannot be excluded from this analysis at $m_f=0$ alone. 

\begin{figure}
 \begin{center}
  \leavevmode
  \epsfxsize=9cm \epsfbox{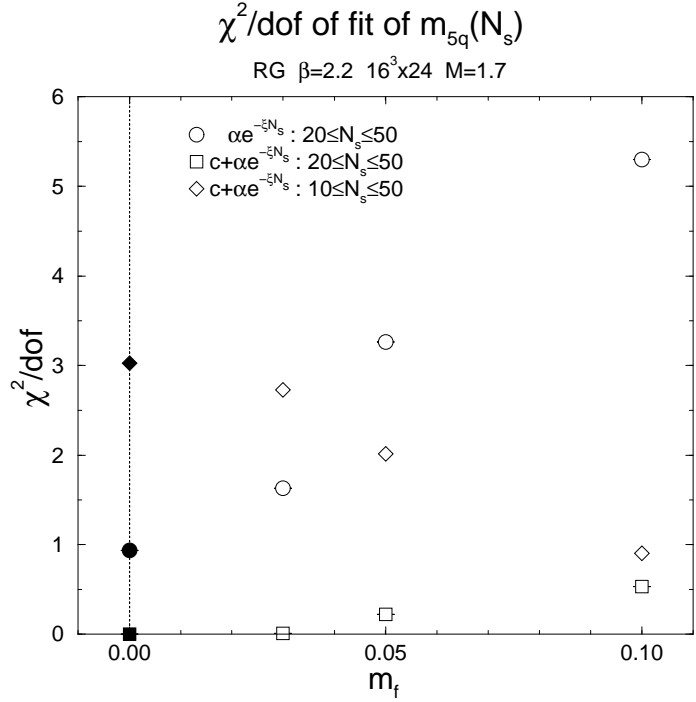}
  \caption{$\chi^2/dof$ in the five and four points fit of $m_{5q}$ data 
  in strong coupling RG action at $\beta=2.2$, $M=1.7$.
  Open circles represent those from the four points fit with simple
  exponential form $\alpha e^{-\xi N_s}$. 
   }
  \label{fig:m5qchi-rg}
 \end{center}
\end{figure}

To further explore this issue, we attempt to fit the data for $m_f\ne 0$. 
We find that 
the plain exponential
does not fit the data at non-zero $m_f$, leading to a large $\chi^2/{\rm dof}$ 
as shown in Fig.~\ref{fig:m5qchi-rg}.
Since we expect the anomalous quark mass to depend little on $m_f$ from
the transfer matrix formalism of DWQCD, this suggests  
that the good fit we obtained without a constant at
$m_f=0$ is perhaps accidental. While further data are needed for 
a definitive conclusion, we think it likely that 
a non-zero anomalous quark mass also remains in the $N_s\to\infty$ limit
at $M=1.7$.

The value of the residual quark mass is small. From the five-point fit, 
we obtain
\begin{eqnarray}
m_{5q}&=&3.60(45) \;{\rm MeV} \quad{\rm at}\  M=1.7\\
      &=&2.87(23) \;{\rm MeV} \quad{\rm at}\  M=2.1
\label{eq:m5q_RG}
\end{eqnarray}

\begin{figure}
 \begin{center}
  \leavevmode
  \epsfxsize=9cm \epsfbox{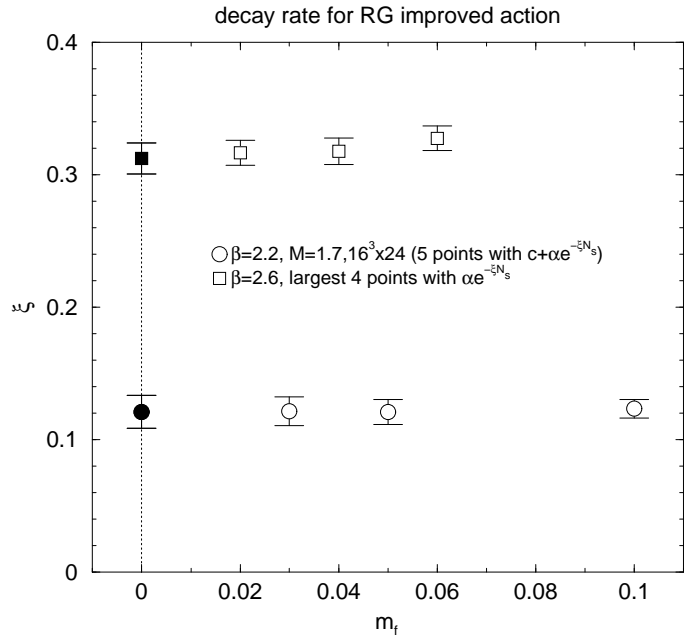}
  \caption{Decay rate $\xi$ of the anomalous quark mass at each $m_f$
  for the RG-improved action.
  Open circles represent data in the strong coupling region
  from the five points fit with $c+\alpha e^{-\xi N_s}$ at $M=1.7$.
  The open squares are those in the weak coupling region from the four
  points fit with $\alpha e^{-\xi N_s}$ at $M=1.8$.}
  \label{fig:drate_mqR5}
 \end{center}
\end{figure}

We finally show the decay rates $\xi$ 
at finite $m_f$ and $m_f=0$ 
in Fig.~\ref{fig:drate_mqR5}.
The decay rates at $\beta=2.2$ are 
obtained by the five-point fit with a constant, while those at 
$\beta=2.6$ employ plain exponential fits
to the four data points.  As expected, the decay rates depend little 
on $m_f$;  their values at $m_f\ne 0$ are consistent with 
those at $m_f=0$. 

We conclude that the RG-improved action significantly improves the chiral 
behavior of DWQCD. 
While chiral symmetry is still not realized at $a^{-1} \simeq 1$ GeV,
the residual anomalous quark mass is sizably reduced compared with
that with the plaquette action.
The results at a weaker coupling ($a^{-1} \simeq 2$~GeV) 
are consistent with the vanishing anomalous quark mass, 
supporting the realization of chiral symmetry with DWQCD.

\section{Discussion}
\label{sec:discussion}


Let us now try to understand our findings. 
The anomalous quark mass $m_{5q}$ vanishes 
exponentially in $N_s$ if the eigenvalues of the transfer matrix 
in the fifth direction are strictly less than 
unity\cite{Shamir95,Neuberger97,HJL,Kikukawa}.
The occurrence of a unit eigenvalue, in turn, is 
in one-to-one correspondence with that of a zero eigenvalue 
of the four dimensional Hermitian Wilson-Dirac 
operator $H_W$\cite{HJL,Kikukawa} or more complicated one, $\hat H_W$,
defined through the fifth dimensional transfer matrix 
$\hat{T}$\cite{Shamir95,Neuberger97},

It has been argued that zero eigenvalues of $H_W$ are related to the existence 
of the parity-flavor broken phases in ordinary lattice QCD with the Wilson 
quark action\cite{Aoki84,SDB,AG,AKU,SS98}.
The connection follows from the identity for the 
parity-flavor order parameter
$\langle \bar q i\gamma_5\tau^3 q \rangle$
given by 
\begin{eqnarray}
\lim_{H\rightarrow+0}\vev{\bar q i\gamma_5\tau^3 q}
&=&
-\lim_{H\rightarrow+0} {\rm Tr}\frac{i\gamma_5\tau^3}{D_W+i\gamma_5\tau^3 H}
=-\lim_{H\rightarrow+0} {\rm tr}\left[
\frac{i\gamma_5}{D_W+i\gamma_5 H} - \frac{i\gamma_5}{D_W-i\gamma_5 H} \right]
\nn \\
&=& -i\lim_{H\rightarrow+0} {\rm tr}
\left[\frac{1}{H_W+i H} - \frac{1}{H_W-i H} \right]
\nn\\
&=& -i\lim_{H\rightarrow+0}\int d\,\lambda \, \rho_{H_W}(\lambda)
\VEV{\lambda}{\left(\frac{1}{\lambda+i H} - \frac{1}{\lambda-i H} 
\right)}{\lambda}
\nn \\
&=& -i \int d\,\lambda \,\rho_{H_W}(\lambda) (-2\pi i)\delta(\lambda) 
= -2\pi\,\rho_{H_W}(0) ,
\label{eq:order}
\end{eqnarray}
where $H$ is an external field coupled to the order parameter, 
$H_W = \gamma_5 D_W$ with $D_W$ the Wilson-Dirac operator,
and $\rho_{H_W}(\lambda)$ is the density of the eigenvalues of $H_W$.

\begin{figure}
 \begin{center}
  \leavevmode
  \epsfxsize=9cm \epsfbox{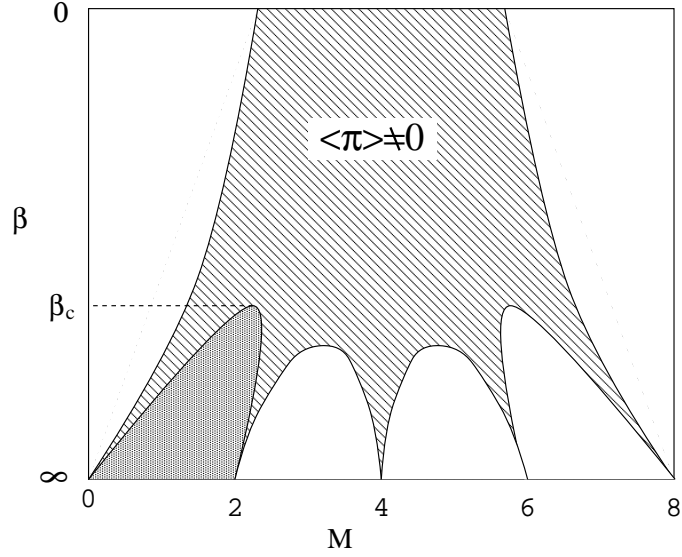}
  \caption{Phase structure of the four dimensional Wilson
  fermion system. $M$ represents the fermion mass (domain-wall height).
  The five cusps at $\beta=\infty$ correspond to the points
  where subsets of $16$ fermions become massless.
  The shaded area marked by $<\pi>\ne 0$ represents the parity broken phase.
  }
  \label{fig:phase}
 \end{center}
\end{figure}

In Fig.\ref{fig:phase}, the expected phase diagram is schematically drawn 
in the $(\beta, M)$ plane, where $-M$ is the bare quark mass of the Wilson
quark action. In the shaded regions the parity-flavor symmetry is broken 
spontaneously corresponding to a non-zero density of zero eigenvalues of 
$H_W$. 
At the critical lines that form the phase boundaries, the neutral pion mass
vanishes. At a strong coupling ($\beta < \beta_c$) there are only two
critical lines, while there exist ten of them at a weak coupling ($\beta >
\beta_c$) and 5 points where two lines meet at $\beta=\infty$.
Each point corresponds to one continuum limit, whose low energy spectra are
composed of a part of sixteen fermion doublers.

For DWQCD to work we have to tune the domain-wall height $M$ within the thick
shaded region in Fig.~\ref{fig:phase} in order to 
avoid zero eigenvalues of $H_W$.  In the weak coupling limit, this 
region is given by $0 < M < 2$.
As the coupling increases, the range is shifted to a larger value of $M$
and shrinks in width.  Finally no massless fermion exists in 
the strong coupling region at $\beta < \beta_c$.

A naive interpretation of our results for DWQCD according to this picture 
would be that $\beta_c > 6.0$ for the plaquette action while
$2.6> \beta_c > 2.2$ for the RG-improved action, or 
in terms of the lattice spacing $a_c^{-1}$ evaluated at $\beta=\beta_c$, 
$a_c^{-1}\gsim 2$~GeV
for the former action and 2~GeV$\gsim a_c^{-1}\gsim 1$~GeV for the latter.  

This interpretation seems inconsistent with the quenched spectrum result 
for the plaquette action obtained in Ref.~\cite{AKU}; in this work,   
non-zero pion mass, signalling the parity conserving phase, 
was reported over the range $0.909(11)\leq M \leq 2.47(5)$ at $\beta=6.0$, 
indicating $\beta_c<6.0$. 
A possible explanation reconciling this apparent conflict is as follows.
If the order parameter, eq.~\eqn{eq:order}, remains non-zero in the
$H\rightarrow+0$ limit, the charged pions become massless Nambu-Goldstone
modes associated with spontaneous parity-flavor breaking in the infinite
volume limit.
The fact that $m_{5q}$ in the proper chiral limit is tiny at $\beta = 6.0$
suggests that the magnitude of the order parameter is
extremely small.
Since the simulation in Ref.~\cite{AKU} is performed on a finite volume
($16^3$) without adding the external field $H$,
it may be difficult to detect the existence of massless pions corresponding to 
such a small order parameter.
The anomalous quark mass, $m_{5q}$,
on the other hand, is very sensitive to the small eigenvalues of $H_W$,
and hence to the non-zero order parameter.
To clarify this issue, a more detailed study for the parity-flavor
breaking phase of ordinary lattice QCD with the Wilson quark will be needed.

The actual dynamics inducing non-zero $\rho_{H_W}(0)$ may be complicated, 
possibly involving instantons.  
The relation between the spectral gap of the Hermitian Wilson-Dirac 
operator $H_W$ and the instanton number has been extensively 
studied\cite{EHN00}.  While a single instanton causes a zero eigenvalue 
of $H_W$ only for a single value of $M$, 
an ensemble of instantons may lead to a non-zero density $\rho_{H_W}(0)\ne 0$. 
In particular unphysical short-distance topological dislocations, having 
an action less than that of physical instantons that are present for 
the plaquette action, may lead to such an effect\cite{EHN00}.  
They will lead to an anomalous quark mass even 
at $a^{-1}\simeq 2$~GeV as we have found in our simulations. 

This mechanism may also explain the difference in the lattice spacing 
needed to realize chiral symmetry in DWQCD between the plaquette and 
RG-improved actions. 
The RG-improved gauge action suppresses unphysical dislocations by 
pushing the action above that of physical
instantons\cite{dislocationRG}. Hence their
influence on the density of zero-eigenvalues will be negligibly small
compared to that for the plaquette action.    
Work exploring zero eigenvalues of the Wilson-Dirac operator for both gauge 
actions\cite{Nagai00} should shed light on this interesting problem. 

\section{Conclusions}
\label{sec:conclusion}


We have investigated the chiral property of domain-wall QCD
within the quenched approximation.  
We have defined the anomalous quark mass (axial WT identity quark mass) 
as an indicator for the realization of chiral symmetry. 
Our simulations have been made in both strong and weak coupling regions
corresponding to $a^{-1}\simeq1$ and
2 GeV, using the plaquette and RG-improved gauge actions.
We have found that the anomalous quark mass remains non-zero 
in the DWQCD chiral limit, $m_f\to0$ and $N_s\to\infty$ 
for the plaquette action for $a^{-1}\le 2$ GeV. 
The magnitude of chiral symmetry breaking rapidly 
decreases with lattice spacing, but the exact chiral symmetry is not
realized at least for this lattice spacing. 

On the contrary, our analysis for the RG-improved 
action reveals a much improved chiral behavior. 
The anomalous quark mass vanishes 
exponentially with this action at 
$a^{-1}\simeq 2$~GeV, indicating that the exact chiral symmetry is realized.  
At $a^{-1}\simeq 1$~GeV the improvement of the chiral behavior 
is not sufficient 
to remove a non-zero anomalous quark mass.


Overall, quenched domain-wall QCD at $a^{-1}\simeq1$ GeV 
appears no better than usual lattice QCD with 
the ordinary Wilson quark action.  
Since the effect of explicit chiral symmetry breaking is
non-negligible, the operators relevant for the electroweak matrix elements
mix nontrivially between different chiralities.
There is no {\it a~priori} reason to expect that these mixing
coefficients are small.

The situation is better in the weak coupling region, 
e.g.,at $a^{-1}\simeq 2~$GeV. 
Even with the plaquette action, the chiral symmetry breaking effect 
is significantly smaller than the physical $u,d$ quark masses, and hence 
may not seriously affect the chiral properties of weak
matrix elements\cite{RBC9909107}.
Moreover, the chiral property of DWQCD is distinctly 
improved with the RG action:
not only the explicit breaking is shown to vanish exponentially in $N_s$,
but also the size of the breaking effect at finite $N_s$ is smaller. 
The improved
gauge actions, such as the RG-improved one, should be a preferred choice 
for future numerical simulations for DWQCD.

\section*{Acknowledgments}

This work is supported in part by the Grants-in-Aid of Ministry of
Education (Nos. 09304029, 10640246, 10640248, 11640250, 10740107,
11640250, 11640294, 11740162, 12640253, 12014202, 2373).
AAK and TM are supported by the JSPS Research for the 
Future Program (No. JSPS-RFTF 97P01102).
SE, KN and JN are JSPS Research Fellows. 

\newcommand{\J}[4]{{#1} {\bf #2} (#3) #4}
\newcommand{\AP}{Ann.~Phys.}
\newcommand{\CMP}{Commun.~Math.~Phys.}
\newcommand{\IJMP}{Int.~J.~Mod.~Phys.}
\newcommand{\MPL}{Mod.~Phys.~Lett.}
\newcommand{\NP}{Nucl.~Phys.}
\newcommand{\NPSup}{Nucl.~Phys.~B (Proc.~Suppl.)}
\newcommand{\PL}{Phys.~Lett.}
\newcommand{\PR}{Phys.~Rev.}
\newcommand{\PRL}{Phys.~Rev.~Lett.}
\newcommand{\PTP}{Prog. Theor. Phys.}
\newcommand{\Suppl}{Prog. Theor. Phys. Suppl.}

\newcommand{\hu}[1]{{\bf #1}}

\begin{table}[p]
\begin{center}
\begin{tabular}{c|ccccc|ccccc}
      & \multicolumn{5}{c|}{plaquette ($\beta=5.65$)} 
        & \multicolumn{5}{c}{RG ($\beta=2.2$)} \\
$N_s$ & 10 & 20 & 30 & 40 & 50 & 10 & 20 & 30 & 40 & 50\\
\hline
$M$   & \multicolumn{10}{c}{$12^3\times 24$} \\
\hline
1.3   & 20 & 20 & 20 & \hu{20} & \hu{20} & 20 & 20 & 20 & \hu{20} & \hu{20}\\
1.7   & 30 & 20 & 20 & \hu{20} & \hu{20} & 30 & 30 & 20 & \hu{40} & \hu{20}\\
2.1   & 30 & 30 & 30 & \hu{20} & \hu{20} & 30 & 30 & 20 & \hu{20} & \hu{20}\\
2.5   & 30 & 20 & 20 & -  & 10 &  30 & 30 & 20 & - &  -\\
\hline
$M$   & \multicolumn{10}{c}{$16^3\times 32$} \\
\hline
1.3   & \hu{20}& \hu{20} & -  & -  & \hu{20}
      & \hu{20} & - & \hu{24} & -  & \hu{24}\\ 
1.7   & \hu{20} & \hu{20} & \hu{20} & \hu{20} & \hu{20} 
      & \hu{20} & \hu{24} & \hu{20} & \hu{20} & \hu{24}\\
2.1   & \hu{20} & \hu{20} & \hu{20} & \hu{20} & \hu{20} 
      & \hu{20} & \hu{20} & \hu{20} & \hu{20} & \hu{24}\\
\end{tabular}
\begin{tabular}{c|ccccc|ccccc}
      & \multicolumn{5}{c|}{plaquette ($\beta=6.00$)} 
        & \multicolumn{5}{c}{RG ($\beta=2.6$)} \\
$N_s$ & 4  & 10 & 16 & 20 & 30 & 4 & 10 & 16 & 20 & 24\\
\hline
\hline
$M$   & \multicolumn{10}{c}{$12^3\times 24$} \\
\hline
1.8   & \hu{20} & \hu{40} & \hu{60} & \hu{80} & \hu{100} 
        &\hu{20} & \hu{40} & \hu{60} & \hu{80} & \hu{100}\\
\hline
\end{tabular}
\end{center}
\caption{
Number of configurations for the plaquette action and the RG
improved action(RG) for each $N_s$. The anomalous quark mass is measured
only at the points where number of configurations is written in {\bf boldface}.
 }
\label{params}
\end{table}

\begin{table}[p]
  \begin{center}
    \leavevmode
    \begin{tabular}{cccccccccc}
     & & fitting range & \multicolumn{4}{c}{$c + \alpha e^{-\xi N_s}$}
     & \multicolumn{3}{c}{$\alpha e^{-\xi N_s}$}\\
     \multicolumn{1}{c}{\raisebox{1.5ex}[0pt]{$\beta$}}
     & \multicolumn{1}{c}{\raisebox{1.5ex}[0pt]{$M$}}
     & of $N_s$ & $c$ & $\alpha$ & $\xi$ & $\chi^2/dof$
     & $\alpha$ & $\xi$ & $\chi^2/dof$\\
      \hline
     & & 10--50 & 0.01032(46) & 0.115(5) & 0.0947(44) & 1.07
     & 0.087(9) & 0.0488(52) & 59.1\\
     & \multicolumn{1}{c}{\raisebox{1.5ex}[0pt]{1.7}} &
        20--50 & 0.00918(107) & 0.08(2) & 0.077(12) & 0.00193
     & 0.051(7) & 0.0337(41) & 9.05\\
      \multicolumn{1}{c}{\raisebox{1.5ex}[0pt]{$5.65$}} & &
        10--50 & 0.00725 (63) & 0.061(3) & 0.0838(69) & 2.23
     & 0.050(4) & 0.0438(46) & 34.8\\
      & \multicolumn{1}{c}{\raisebox{1.5ex}[0pt]{2.1}} &
        20--50 & 0.00524 (149) & 0.040(6) & 0.056(13) & 0.202
     & 0.034(2) & 0.0313(21) & 2.27\\
      \hline
     & $1.8$ & \multicolumn{1}{c}{\raisebox{0ex}[0pt]{4--30}}
     & 0.000565 (147) & 0.12 (1) & 0.364 (22) & 8.70
     & 0.11 (2) & 0.333 (38) & 47.8\\
     \multicolumn{1}{c}{\raisebox{1.5ex}[0pt]{$6.0$}}
     & $1.8$ & \multicolumn{1}{c}{\raisebox{0ex}[0pt]{10--30}}
     & 0.000396 (79) & 0.023 (6) & 0.195 (25) & 0.162
     & 0.012 (3) & 0.123 (19) & 7.29\\
    \end{tabular}
  \caption{Exponential fit of the anomalous quark mass $m_{5q}$ for the
   plaquette action.
   Strong coupling results at $\beta=5.65$ are for the lattice size 
   $16^3\times 24$, 
   and those at a weak coupling of $\beta=6.0$ are for $16^3\times 32$ 
   lattice. 
   Fits with all of the five date points are represented by the fitting
   range 10--50 and 4--30.
   Four-points fits without $N_s=10$ data are represented by 20--50
   and 10--30.}
    \label{tab:m5qfit-pl}
  \end{center}
\end{table}

\begin{table}[p]
 \begin{center}
  \leavevmode
  \begin{tabular}{cccccccccc}
   & & fitting range & \multicolumn{4}{c}{$c + \alpha e^{-\xi N_s}$}
   & \multicolumn{3}{c}{$\alpha e^{-\xi N_s}$}\\
   \multicolumn{1}{c}{\raisebox{1.5ex}[0pt]{$\beta$}}
   & \multicolumn{1}{c}{\raisebox{1.5ex}[0pt]{$M$}} & of $N_s$
   & $c$ & $\alpha$ & $\xi$ & $\chi^2/dof$
   & $\alpha$ & $\xi$ & $\chi^2/dof$\\
   \hline
   & & 10--50 & 0.00371(46) & 0.071(9) & 0.121(12) &  3.03
   & 0.045(9) & 0.066(11) & 35.7\\
   & \multicolumn{1}{c}{\raisebox{1.5ex}[0pt]{1.7}} &
        20--50 & 0.00230(107) & 0.025(7) & 0.0620(201) & 0.00002
   & 0.020(2) & 0.0370(28) & 0.936\\
      \multicolumn{1}{c}{\raisebox{1.5ex}[0pt]{$2.2$}} & &
        10--50 & 0.00296 (24) & 0.036(3) & 0.0979(74) & 0.232
   & 0.025(4) & 0.0505(76) & 21.0\\
      & \multicolumn{1}{c}{\raisebox{1.5ex}[0pt]{2.1}} &
        20--50 & 0.00278 (43) & 0.03(1) & 0.085(21) & 0.113
   & 0.014(2) & 0.0328(51) & 3.98\\
     \hline
   & $1.8$ & \multicolumn{1}{c}{\raisebox{0ex}[0pt]{4--24}}
   & 0.0000099 (71) & 0.122(9) & 0.501(15) & 12.8
   & 0.12(1) & 0.498(17) & 16.9\\
     \multicolumn{1}{c}{\raisebox{1.5ex}[0pt]{$2.6$}}
   & $1.8$ & \multicolumn{1}{c}{\raisebox{0ex}[0pt]{10--24}}
   & $-0.0000071(63)$ & 0.014(3) & 0.281(25) & 0.00227
   & 0.019(2) & 0.312 (12) & 0.799\\
  \end{tabular}
  \caption{Exponential fit of the anomalous quark mass $m_{5q}$ for the
   RG-improved action.
   Strong coupling results at $\beta=2.2$ are for the lattice size 
   $16^3\times 24$, 
   and those at a weak coupling of $\beta=2.6$ are for $16^3\times 32$ 
   lattice. 
   Fits with all of the five date points are represented by the fitting
   range 10--50 and 4--30.
   Four-points fit without $N_s=10$ data are represented by 20--50
   and 10--24.}
  \label{tab:m5qfit-rg}
 \end{center}
\end{table}

\section*{Appendix A: Pion mass}
\label{sec:pion}

The pion mass is a standard quantity in examining the chiral property,
and hence often studied in the context of DWQCD.   
Analyses of this observable, however, is not straightforward due to 
possible finite size effects, particularly for small pion masses, 
and difficulties associated with the chiral extrapolation 
because of chiral logarithms, which may be quite significant in quenched QCD. 
Numerically the pion mass has to be extracted from an exponential falloff 
of the pion propagator.  Even though the pion propagator has the best 
statistical quality among hadron propagators, this procedure is more
susceptible to statistical and systematic uncertainties than the 
determination of the anomalous quark mass which involves only a constant fit 
to a ratio of two kinds of propagators.  For these reasons we 
have employed the anomalous quark mass in the main body of the 
present paper.
In this appendix we present our results for the pion mass, and discuss 
to what extent they match with those obtained with the anomalous 
quark mass.

\begin{figure}[p]
 \begin{center}
  \leavevmode
  \epsfxsize=7cm \epsfbox{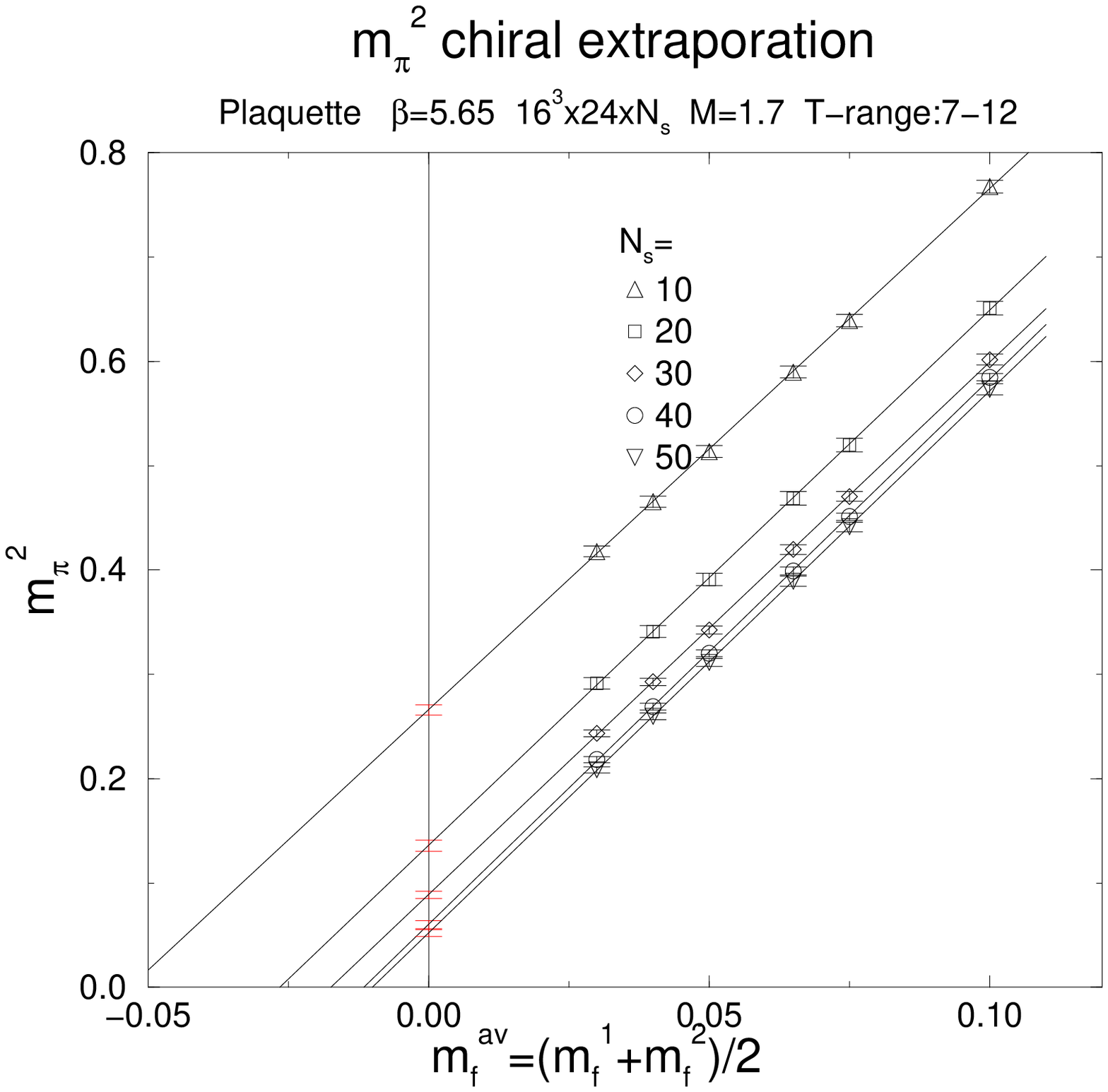}
  \epsfxsize=7cm \epsfbox{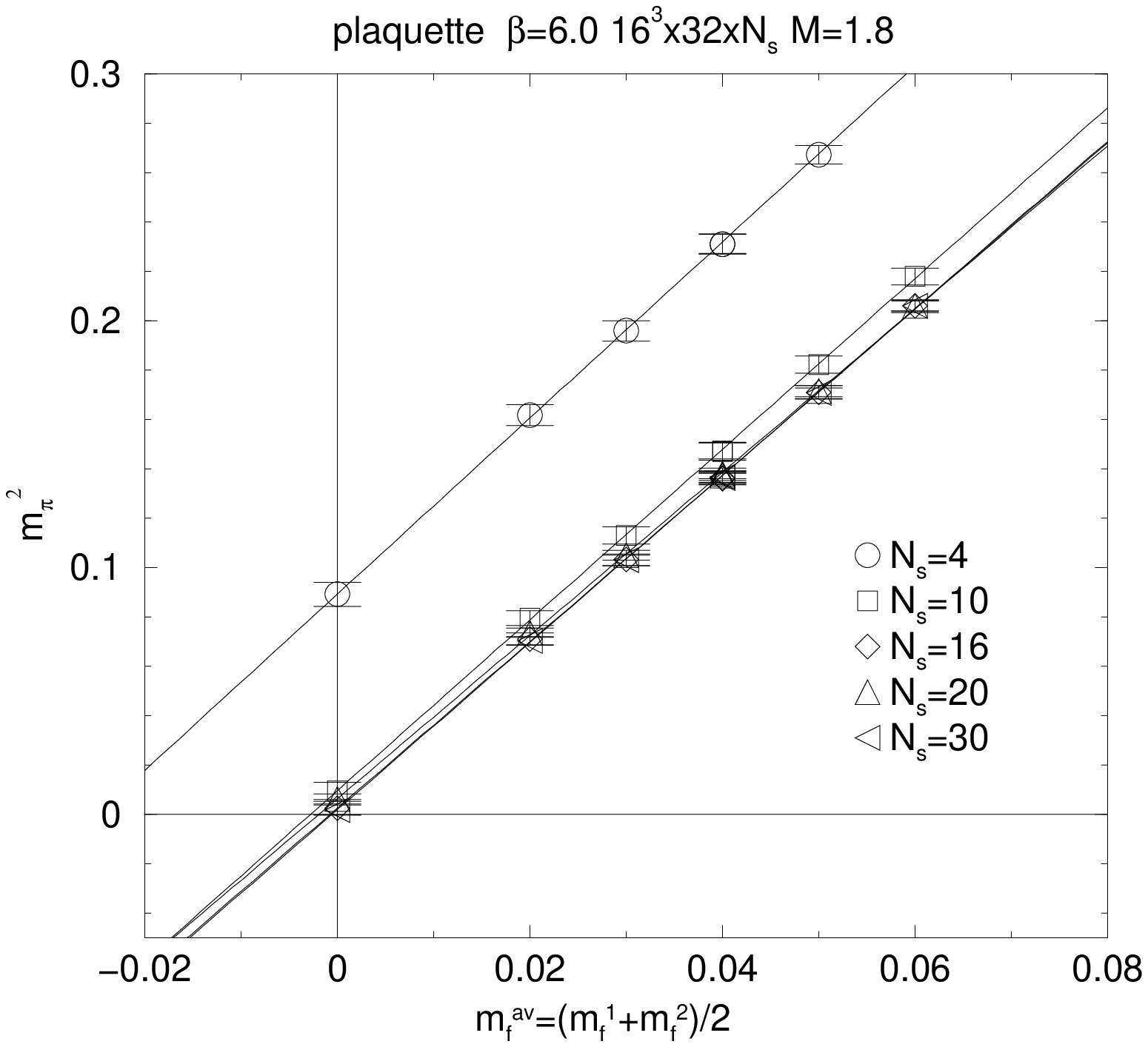}
  \caption{Pion mass squared as a function of valence quark mass $m_f$
  at $M=1.7, \beta=5.65$ on  $16^3\times24\times N_s$ lattice
  (left) and $\beta=6.0$ on $16^3\times32\times N_s$ lattice (right)
  of plaquette action.
  Lines are linear fits to those data.}
    \label{fig:pi2-mf}
 \end{center}
\end{figure}

Typical results for pion mass squared $m_\pi^2(m_f,N_s,M)$ are plotted as a
function of the averaged bare quark mass $m_f^{av}=(m_{f1}+m_{f2})/2$ 
for the plaquette action in Fig.~\ref{fig:pi2-mf}.
The result shows a good linear behavior in $m_f^{av}$. 
The pion mass in the chiral limit $m_f\to0$ is therefore estimated 
by linearly extrapolating $m_\pi^2$ in $m_f^{av}$.
The numerical values of the pion mass are collected in
Appendix B.

\subsection*{Strong coupling region}

\begin{figure}[p]
 \begin{center}
  \leavevmode
  \epsfxsize=9cm \epsfbox{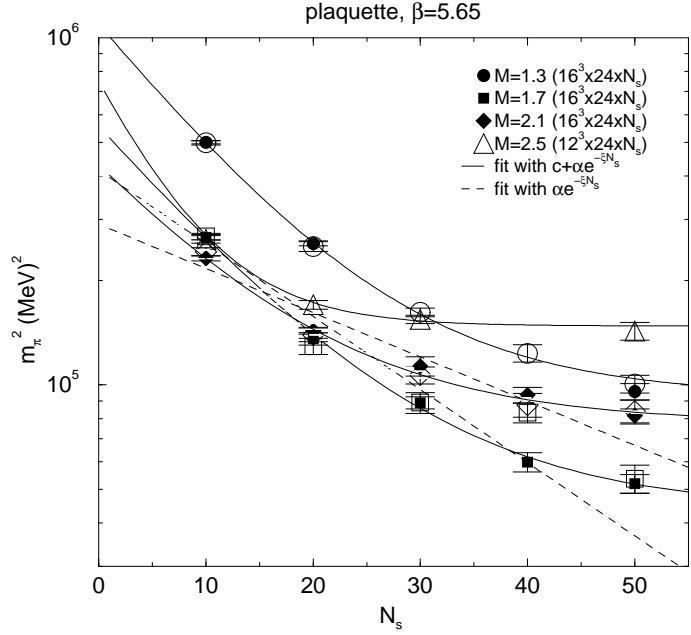}
  \caption{Pion mass squared at $m_f=0$ as a function of the extra
  dimension $N_s$ in the strong coupling region $\beta=5.65$ for 
  plaquette action.  Filled symbols are taken on 
  $16^3 \times 24 \times N_s$ lattice, and open ones on 
  $12^3 \times 24 \times N_s$ lattice. 
 Lines are fits to an exponential with a constant of all five 
  data points on an $N_\sigma=16$ lattice except for 
  $M=1.3$ and $2.5$ for which data fitted are for $N_\sigma=12$.}
   \label{fig:pi20-NsP-5}
 \end{center}
\end{figure}

\begin{figure}[p]
 \begin{center}
  \leavevmode
  \epsfxsize=9cm \epsfbox{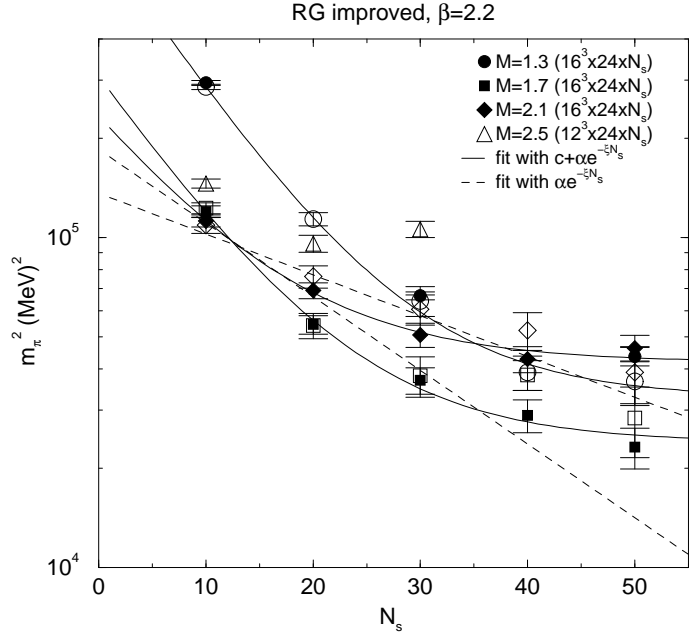}
  \caption{Pion mass squared at $m_f=0$ as a function of the extra
  dimension $N_s$ in the strong coupling region $\beta=2.2$ for 
  RG-improved action.  Filled symbols are taken on 
  $16^3 \times 24 \times N_s$ lattice, and open ones on 
  $12^3 \times 24 \times N_s$ lattice. 
  Lines are fits to a constant and an exponential of all five 
  data points on an $N_\sigma=16$ lattice except for 
  $M=1.3$ for which data fitted are for $N_\sigma=12$.}
  \label{fig:pi20-NsRG-5}
 \end{center}
\end{figure}

We first discuss pion mass results 
at the strong coupling $a^{-1}\simeq 1$ GeV.
In Fig.~\ref{fig:pi20-NsP-5} 
we plot the pion mass squared $m_\pi^2$ in the chiral limit $m_f^{av}=0$
as a function of $N_s$ for the plaquette action at $\beta=5.65$. 
Filled symbols are obtained on a lattice of spatial size $N_\sigma=16$ 
and open ones with a spatial size of $N_\sigma=12$. 
A similar figure for the RG-improved action is shown in 
Fig.~\ref{fig:pi20-NsRG-5}.  

Solid lines in these figures show results of fits by a form 
$c+\alpha e^{-\xi N_s}$ using all five points obtained on an $N_\sigma=16$ 
lattice ($M=1.7,2.1$) or an $N_\sigma=12$ lattice ($M=1.3,2.5$). 
They reproduce the data well for a non-zero $c$ with an acceptable
$\chi^2$ as summarized in Tables~\ref{tab:mpifit-16} and
\ref{tab:mpifit-12}.
Therefore pion mass does not vanish in the chiral limit
$m_f\to0$ and $N_s\to\infty$ at $a^{-1}\simeq 1$~GeV for the plaquette
and the RG-improved actions, which is consistent with 
the conclusions from the analysis of the anomalous quark mass.

\begin{figure}[p]
 \begin{center}
  \leavevmode
  \epsfxsize=7cm \epsfbox{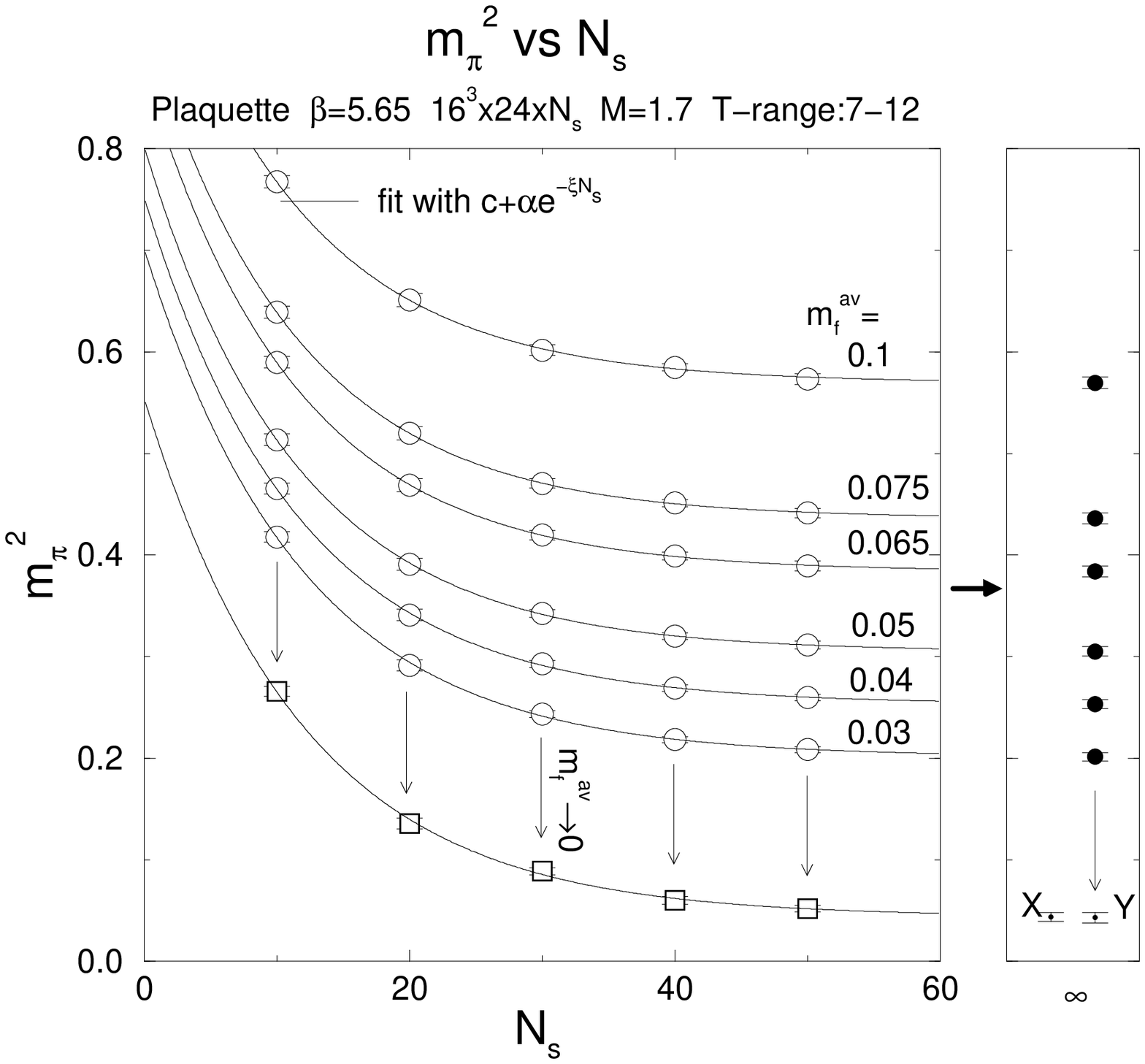}
  \epsfxsize=7cm \epsfbox{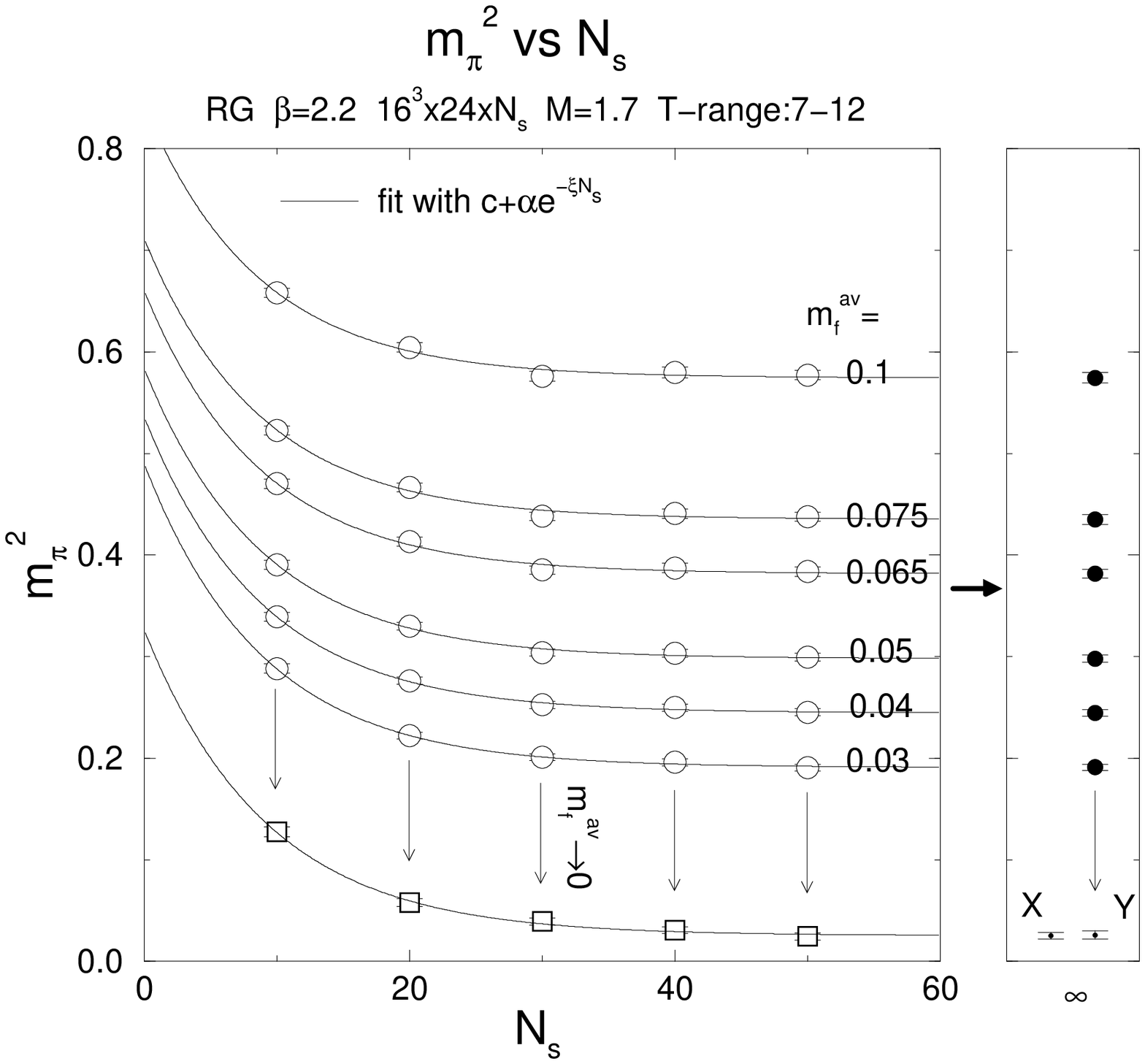}
  \caption{Pion mass squared as a function of the extra dimension $N_s$
  at $M=1.7$ on $16^3 \times 24 \times N_s$ lattice.
  The left figure represents the results for the plaquette action at
  $\beta=5.65$.
  The right figure is for the RG action at $\beta=2.2$.
  The pion masses derived with two different order of limits $m_f\to0$,
  $N_s\to\infty$ are represented with ``X'' and ``Y''.}
  \label{fig:pi2-NsP}
 \end{center}
\end{figure}

In order to check our procedure of taking the chiral limit we
interchange the order of the limits $m_f^{av}\to 0$ and
$N_s\to\infty$.
As shown by solid lines going through open circles 
in  Fig.~\ref{fig:pi2-NsP}, we first make a fit of form
$m_\pi^2(m_f^{av},N_s) = c^\prime(m_f^{av})+\alpha e^{-\xi N_s}$ 
for each value of $m_f^{av}$.
We note that the constant term has to be included in the fit of the
pion mass at non-zero $m_f$.
A linear chiral extrapolation 
$m_\pi^2(m_f^{av},N_s=\infty)=c^\prime(m_f^{av})=d+\gamma m_f^{av}$
then yields $d = 0.0432(52)$, which agrees well with the value of
$c=0.0440(43)$ previously obtained.
The commutativity of the two limits is summarized by two symbols
``X'' and ``Y'' in Fig.~\ref{fig:pi2-NsP},
where X represents $m_\pi^2(m_f^{av}=0, N_s=\infty)$ given with
$m_f\to0$, then $N_s\to\infty$ and Y represents the pion mass with the
limit $N_s\to\infty$, then $m_f\to0$.
These two values agree very well with each other.
The same consistency check is also made for the RG-improved action.

\begin{figure}[p]
 \begin{center}
  \leavevmode
  \epsfxsize=9cm \epsfbox{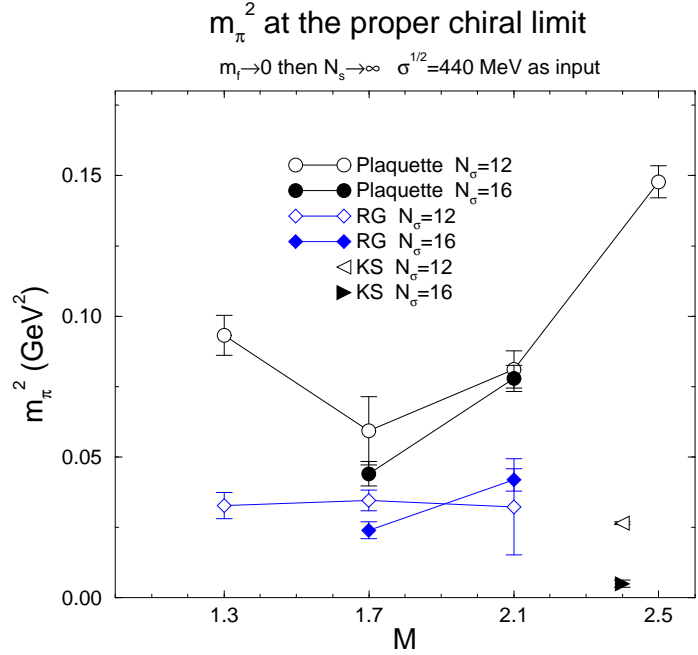}
  \caption{Pion mass squared as a function of domain-wall height $M$
  in the  chiral limit $m_f\to0$, $N_s\to\infty$.
  The results on $16^3\times24$ lattice (open symbols) are compared with
  those on $12^3\times24$ (filled symbols).
  Open and filled triangles are those from the KS fermion action at
  $N_\sigma=12$ and $N_\sigma=16$ lattice volumes, which represent the 
  pure finite volume effect.}
  \label{fig:pi2-M}
 \end{center}
\end{figure}

The above analyses strongly support the conclusion that $m_\pi$ does not
vanish in the limit $m_f^{av}\to0$ and $ N_s\to\infty$.
In order to examine finite spatial volume effects as a possible origin 
of this non-zero mass, we compare the pion mass squared in the proper 
chiral limit $m_\pi^2(m_f^{av}\to0, N_s\to\infty)$ 
for two kinds of lattice volumes $12^3\times24$ (open symbols) 
and $16^3\times24$ (filled symbols) in Fig.~\ref{fig:pi2-M}. 

The circles and diamonds represent the results from the plaquette and the RG
action, respectively.
The horizontal triangles are results for 
the Nambu-Goldstone pion mass of the Kogut-Susskind quark action 
at $\beta=5.7$ for a spatial lattice size of $N_\sigma=12$ 
(open triangle) and $N_\sigma=16$ (filled triangle)\cite{GGKS}. From 
comparison of these data points, we consider that there may be some finite
size effects at $M=1.7$, but that they are not large enough to explain 
non-zero values of pion mass for DWQCD at strong coupling. 

Since there still remains a possibility that this non-zero mass is
caused by the quenching effect,
we adopt the WT identity mass as a better indicator 
to test the chiral property of DWQCD in the main body of this article.

\subsection*{Weak coupling region}

\begin{figure}
 \begin{center}
  \leavevmode
  \epsfxsize=9cm \epsfbox{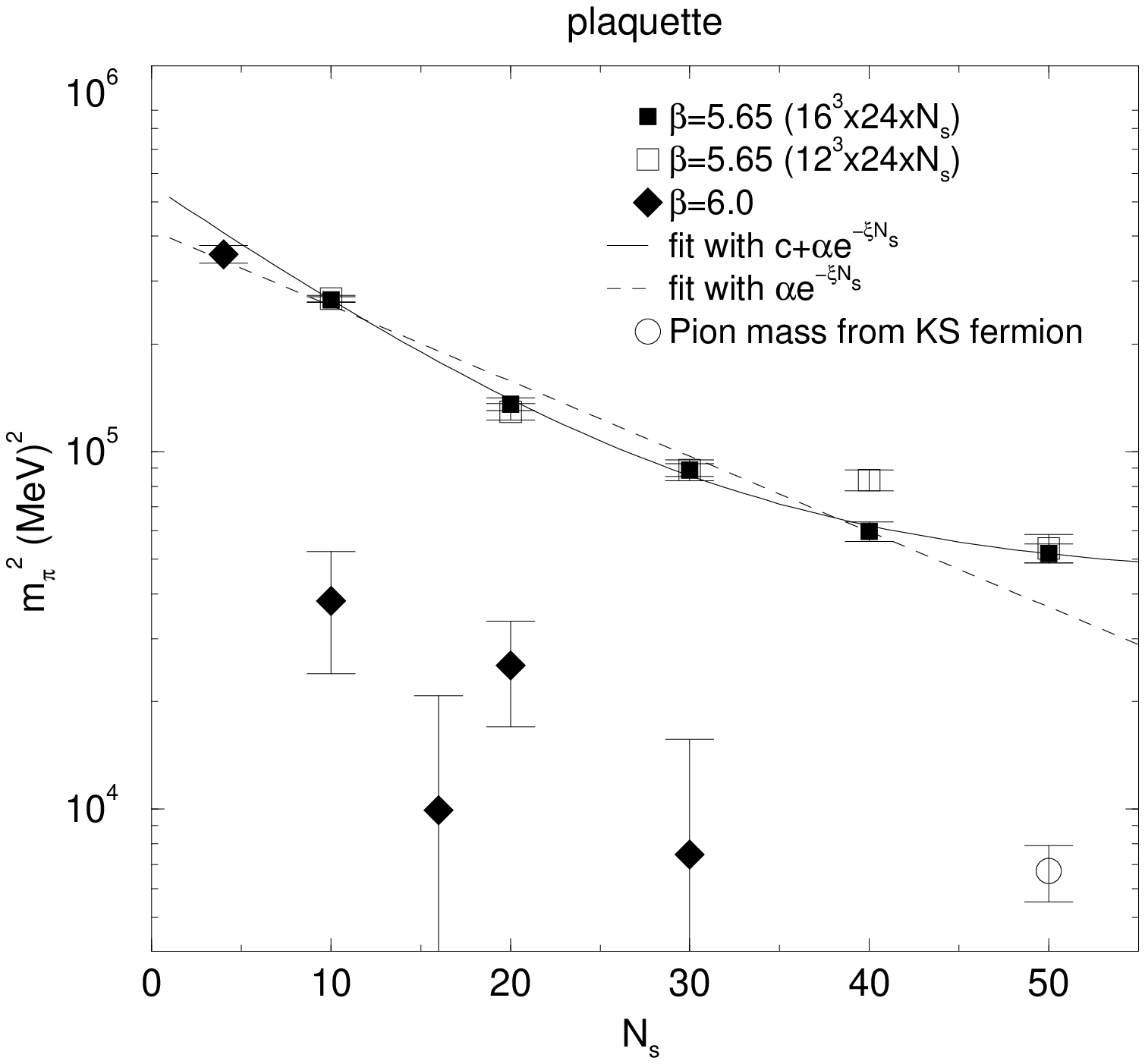}
  \caption{Pion mass squared at $m_f=0$ as a function of $N_s$ at
  $\beta=6.0$ of plaquette action with $M=1.8$ (filled diamonds) as
  compared with those at $\beta=5.65$ and and $M=1.7$ (squares). 
  Open circle represents the pion mass from the KS fermion at
  $N_\sigma=16$ lattice volume.}
  \label{mpi2vNs-P-6.0}
 \end{center}
\end{figure}

\begin{figure}
 \begin{center}
  \leavevmode
  \epsfxsize=9cm \epsfbox{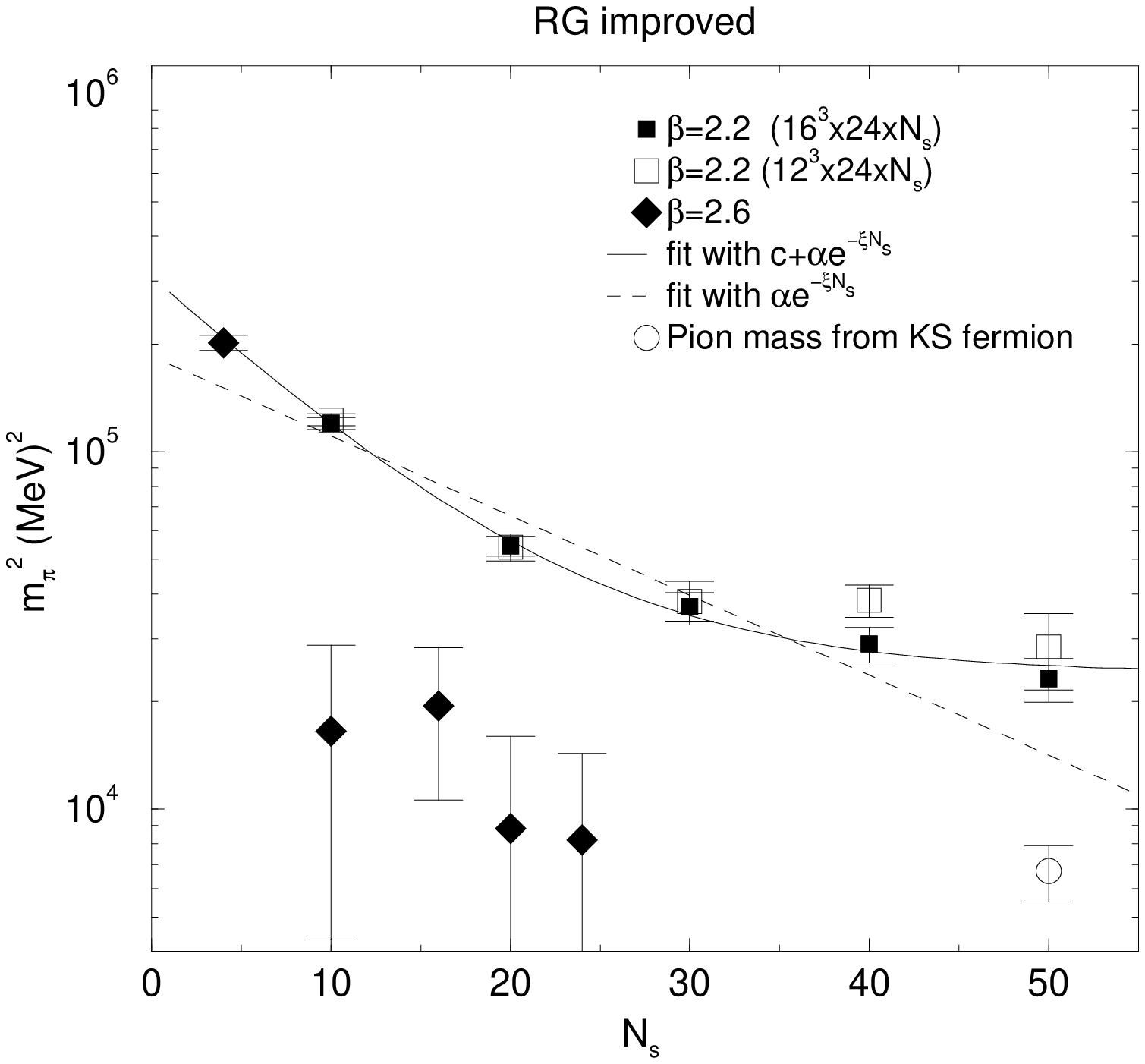}
  \caption{Pion mass squared at $m_f=0$ as a function of $N_s$ at
  $\beta=2.6$ of RG-improved action with $M=1.8$ (filled diamonds) as
  compared with those at $\beta=2.2$ and and $M=1.7$ (squares).
  Open circle represents the pion mass from the KS fermion at
  $N_\sigma=16$ lattice volume.}
  \label{mpi2vNs-RG-2.6}
 \end{center}
\end{figure}

Pion mass squared in the chiral limit $m_\pi^2(m_f=0,N_s,M)$ is
plotted in Fig.~\ref{mpi2vNs-P-6.0} for the plaquette action at $\beta=6.0$ 
and in Fig.~\ref{mpi2vNs-RG-2.6} for the RG-improved action at $\beta=2.6$. 
The results have large errors, and do not show a clear trend as a function 
of $N_s$.  However, the magnitude beyond $N_s=10$ are comparable to 
the value obtained for the Nambu-Goldstone pion mass of
the Kogut-Susskind fermion at $\beta=6.0$ for the same spatial
lattice size $16^3\times40$ (open circle)\cite{GGKS,AU94}. 
This indicates that these non-zero pion masses are mainly caused by finite
spatial volume effects.
Hence the chiral property of DWQCD cannot be studied by pion
mass in the weak coupling region on this volume.

\begin{table}[p]
 \begin{center}
  \leavevmode
  \begin{tabular}{cccccccccc}
   & & fitting range & \multicolumn{4}{c}{$c + \alpha e^{-\xi N_s}$} &
   \multicolumn{3}{c}{$\alpha e^{-\xi N_s}$}\\
   \multicolumn{1}{c}{\raisebox{1.5ex}[0pt]{action}}
   & \multicolumn{1}{c}{\raisebox{1.5ex}[0pt]{$M$}} & of $N_s$
   & $c$ & $\alpha$ & $\xi$ & $\chi^2/dof$
   & $\alpha$ & $\xi$ & $\chi^2/dof$\\
   \hline
     & &
   10--50 & 0.04404(434) & 0.51(3) & 0.0836(62) & 0.838
   & 0.41(4) & 0.0485(54) & 16.1\\
     & \multicolumn{1}{c}{\raisebox{1.5ex}[0pt]{1.7}} &
   20--50 & 0.03946(829) & 0.4(1) & 0.071(16) & 0.944
   & 0.26(4) & 0.0341(46) & 3.63\\
     \multicolumn{1}{c}{\raisebox{1.5ex}[0pt]{plaq.}} & &
   10--50 & 0.07796(464) & 0.35(3) & 0.0832(78) & 0.761
   & 0.29(3) & 0.0293(54) & 20.4\\
     & \multicolumn{1}{c}{\raisebox{1.5ex}[0pt]{2.1}} &
   20--50 & 0.0602(262) & 0.21(5) & 0.046(28) & 0.00354
   & 0.21(1) & 0.0195(16) & 0.523\\
     \hline
     & &
   10--50 & 0.02549(321) & 0.30(4) & 0.109(14) & 0.585
   & 0.20(4) & 0.0513(92) & 8.65\\
     & \multicolumn{1}{c}{\raisebox{1.5ex}[0pt]{1.7}} &
   20--50 & 0.0191(105) & 0.14(8) & 0.063(38) &  0.0661
   & 0.10(1) & 0.0300(46) & 0.456\\
     \multicolumn{1}{c}{\raisebox{1.5ex}[0pt]{RG}} & &
   10--50 & 0.04453(426) & 0.20(4) & 0.100(21) & 0.553
   & 0.14(2) & 0.0285(66) & 7.17\\
     & \multicolumn{1}{c}{\raisebox{1.5ex}[0pt]{2.1}} &
   20--50 & 0.04669(449) & 0.5(1.1) & 0.15(11) & 0.677
   & 0.10(2) & 0.0167(59) & 2.80\\
  \end{tabular}
  \caption{Exponential fit of pion mass squared $m_\pi^2$ in strong
  coupling region at $M=1.7$ and $2.1$.
  The four-dimensional lattice size is $16^3\times24$.
  Fits with all of five data points are represented by the fitting
  range $10-50$ and four-points fits without $N_s=10$ data are 
  represented by $20-50$.}
  \label{tab:mpifit-16}
 \end{center}
\end{table}

\begin{table}[p]
 \begin{center}
  \leavevmode
  \begin{tabular}{cccccccccc}
   & & fitting range & \multicolumn{4}{c}{$c + \alpha e^{-\xi N_s}$} &
   \multicolumn{3}{c}{$\alpha e^{-\xi N_s}$}\\
   \multicolumn{1}{c}{\raisebox{1.5ex}[0pt]{action}}
   & \multicolumn{1}{c}{\raisebox{1.5ex}[0pt]{$M$}} & of $N_s$
   & $c$ & $\alpha$ & $\xi$ & $\chi^2/dof$
   & $\alpha$ & $\xi$ & $\chi^2/dof$\\
   \hline
     & &
   10--50 & 0.09328(709) & 1.00(5) & 0.0904(57) & 0.702
   & 0.8(1) & 0.0500(64) & 27.0\\
     & \multicolumn{1}{c}{\raisebox{1.5ex}[0pt]{1.3}} &
   20--50 & 0.0842(126) & 0.7(2) & 0.075(14) & 0.147
   & 0.45(7) & 0.0320(50) & 5.19\\
     & &
   10--50 & 0.0593(121) & 0.6(1) & 0.100(25) & 5.27
   & 0.40(6) & 0.0454(74) & 16.6\\
     & \multicolumn{1}{c}{\raisebox{1.5ex}[0pt]{1.7}} &
   20--50 & $-0.021(281)$ & 0.2(2) & 0.021(56) & 4.90
   & 0.22(3) & 0.0267(49) & 2.47\\
     \multicolumn{1}{c}{\raisebox{1.5ex}[0pt]{plaq.}}
     & &
   10--50 & 0.08112(660) & 0.43(6) & 0.098(14) & 0.226
   & 0.31(4) & 0.0332(65) & 11.7\\
     & \multicolumn{1}{c}{\raisebox{1.5ex}[0pt]{2.1}} &
   20--50 & 0.0765(140) & 0.3(2) & 0.077(42) & 0.192
   & 0.20(2) & 0.0189(35) & 1.46\\
     & &
   10--50 & 0.14777(566) & 0.6(2) & 0.159(33) & 0.617
   & 0.29(6) & 0.0205(86) & 30.2\\
     & \multicolumn{1}{c}{\raisebox{1.5ex}[0pt]{2.5}} &
   20--50 &  &  & \\
     \hline
     & &
   10--50 & 0.03482(494) & 0.83(7) & 0.1125(84) & 0.437
   & 0.63(9) & 0.0757(98) & 11.4\\
     & \multicolumn{1}{c}{\raisebox{1.5ex}[0pt]{1.3}} &
   20--50 & 0.03175(973) & 0.6(4) & 0.098(34) & 0.670
   & 0.31(6) & 0.0475(69) & 2.23\\
     & &
   10--50 & 0.03681(390) & 0.4(1) & 0.150(31) & 0.704
   & 0.20(4) & 0.048(11) & 9.15\\
     \multicolumn{1}{c}{\raisebox{1.5ex}[0pt]{RG}}
   & \multicolumn{1}{c}{\raisebox{1.5ex}[0pt]{1.7}} &
   20--50 & 0.0327(152) & 0.1(2) & 0.08(12) & 1.30
   & 0.08(2) & 0.0189(60) & 0.759\\
     & &
   10--50 & 0.0344(181) & 0.13(2) & 0.051(25) & 0.281
   & 0.15(1) & 0.0262(35) & 0.529\\
     & \multicolumn{1}{c}{\raisebox{1.5ex}[0pt]{2.1}} &
   20--50 & $-0.035$(239) & 0.2(2) & 0.013(35) & 0.1593
   & 0.12(2) & 0.0209(56) & 0.0883\\
  \end{tabular}
  \caption{Exponential fit of the pion mass squared $m_\pi^2$ in strong
  coupling region at $M=1.3, 1.7, 2.1, 2.5$.
  The four dimensional lattice size is $12^3\times24$.
  Fits with all of five data points are represented by the fitting
  range $10-50$ and four-points fit without $N_s=10$ data are 
  represented by $20-50$.}
  \label{tab:mpifit-12}
 \end{center}
\end{table}

\newpage
\section*{Appendix B}

In this appendix we lists results for anomalous quark mass $m_{5q}$ 
and pion mass squared $m_\pi^2$.
 
\begin{table}[p]
  \leavevmode
\begin{center}
\begin{tabular}{l|lllll}
  $m_f$ & \multicolumn{5}{c}{$N_s$} \\
  & 10 & 20 & 30 & 40 & 50\\
\hline 
& \multicolumn{5}{c}{$\beta=5.65$, $M=1.3$ on $12^3\times24$}\\
\hline
$0.00$ & $-$ & $-$ & $-$ & $0.0605$ ($31$) & $0.0499$ ($20$)\cr
$0.03$ & $-$ & $-$ & $-$ & $0.0622$ ($28$) & $0.0520$ ($19$)\cr
$0.05$ & $-$ & $-$ & $-$ & $0.0629$ ($26$) & $0.0530$ ($19$)\cr
$0.1$ & $-$ & $-$ & $-$ & $0.0658$ ($24$) & $0.0565$ ($21$)\cr
\hline 
& \multicolumn{5}{c}{$\beta=5.65$, $M=1.7$ on $12^3\times24$}\\
\hline
$0.00$ & $-$ & $-$ & $-$ & $0.01303$ ($56$) & $0.01019$ ($47$)\cr
$0.03$ & $-$ & $-$ & $-$ & $0.01304$ ($53$) & $0.01019$ ($42$)\cr
$0.05$ & $-$ & $-$ & $-$ & $0.01302$ ($59$) & $0.01016$ ($42$)\cr
$0.1$ & $-$ & $-$ & $-$ & $0.01302$ ($63$) & $0.01018$ ($44$)\cr
\hline
& \multicolumn{5}{c}{$\beta=5.65$, $M=2.1$ on $12^3\times24$}\\
\hline
$0.00$ & $-$ & $-$ & $-$ & $0.00934$ ($28$) & $0.00770$ ($42$)\cr
$0.03$ & $-$ & $-$ & $-$ & $0.01031$ ($31$) & $0.00894$ ($29$)\cr
$0.05$ & $-$ & $-$ & $-$ & $0.01051$ ($35$) & $0.00907$ ($33$)\cr
$0.1$ & $-$ & $-$ & $-$ & $0.01230$ ($55$) & $0.01155$ ($64$)\cr
\end{tabular}
\caption{Anomalous quark mass for the plaquette action in the
 strong coupling region at $\beta=5.65$, $M=1.3,1.7,2.1$ on $12^3\times24$ 
lattice.}
\label{tbl:anoq-pl-st12M13}
\end{center}
\end{table}

\begin{table}[p]
  \leavevmode
\begin{center}
\begin{tabular}{l|lllll}
  $m_f$ & \multicolumn{5}{c}{$N_s$} \\
   & 10 & 20 & 30 & 40 & 50\\
\hline 
& \multicolumn{5}{c}{$\beta=5.65$, $M=1.3$ on $16^3\times24$}\\
\hline
$0.00$ & $0.1747$ ($16$) & $0.1129$ ($14$) & $-$ & $-$ & $0.04925$ ($98$)\cr
$0.03$ & $0.1723$ ($16$) & $0.1123$ ($16$) & $-$ & $-$ & $0.0513$ ($11$)\cr
$0.05$ & $0.1700$ ($17$) & $0.1111$ ($17$) & $-$ & $-$ & $0.0522$ ($11$)\cr
$0.1$  & $0.1660$ ($18$) & $0.1102$ ($20$) & $-$ & $-$ & $0.0557$ ($14$)\cr
\hline
& \multicolumn{5}{c}{$\beta=5.65$, $M=1.7$ on $16^3\times24$}\\
\hline
$0.00$ & $0.05514$ ($49$) & $0.02717$ ($62$) & $0.01748$ ($55$)
 & $0.01303$ ($27$) & $0.01095$ ($47$)\cr
$0.03$ & $0.05373$ ($46$) & $0.02646$ ($58$) & $0.01721$ ($51$)
 & $0.01280$ ($29$) & $0.01092$ ($39$)\cr
$0.05$ & $0.05228$ ($49$) & $0.02561$ ($59$) & $0.01691$ ($51$)
 & $0.01252$ ($35$) & $0.01084$ ($43$)\cr
$0.1$ & $0.05002$ ($67$) & $0.02445$ ($60$) & $0.01645$ ($54$)
 & $0.01219$ ($52$) & $0.01080$ ($44$)\cr
\hline
& \multicolumn{5}{c}{$\beta=5.65$, $M=2.1$ on $16^3\times24$}\\
\hline
$0.00$ & $0.03349$ ($31$) & $0.01820$ ($37$) & $0.01253$ ($39$)
 & $0.00954$ ($24$) & $0.00757$ ($43$)\cr
$0.03$ & $0.03338$ ($27$) & $0.01941$ ($35$) & $0.01358$ ($37$)
 & $0.01032$ ($22$) & $0.00880$ ($38$)\cr
$0.05$ & $0.03261$ ($28$) & $0.01947$ ($38$) & $0.01361$ ($37$)
 & $0.01032$ ($24$) & $0.00909$ ($46$)\cr
$0.1$  & $0.03254$ ($37$) & $0.02171$ ($56$) & $0.01542$ ($44$)
 & $0.01203$ ($56$) & $0.01124$ ($51$)\cr
\end{tabular}
\caption{Anomalous quark mass for the plaquette action in the
 strong coupling region at $\beta=5.65$, $M=1.3,1.7,2.1$ on $16^3\times24$ 
lattice.}
\label{tbl:anoq-pl-st16M13}
\end{center}
\end{table}

\begin{table}[p]
  \leavevmode
\begin{center}
\begin{tabular}{l|lllll}
$m_f$  & \multicolumn{5}{c}{$N_s$} \\
  & 4 & 10 & 16 & 20 & 30\\
\hline 
$0.0$ & $0.02955$ ($29$) & $0.00373$ ($13$) & $0.00148$ ($17$)
 & $0.00084$ ($11$) & $0.000467$ ($57$)\cr
$0.02$ & $0.02884$ ($25$) & $0.00360$ ($11$) & $0.00134$ ($14$)
 & $0.000750$ ($90$) & $0.000403$ ($49$)\cr
$0.04$ & $0.02804$  ($20$) & $0.003446$ ($78$) & $0.001128$ ($93$)
 & $0.000637$ ($73$) & $0.000317$ ($42$)\cr
$0.06$ & $0.02735$  ($18$) & $0.003316$ ($64$) & $0.000989$ ($66$)
 & $0.000548$ ($59$) & $0.000259$ ($38$)\cr
\end{tabular}
 \caption{Anomalous quark mass for the plaquette action in the
 weak coupling region at
 $\beta=6.0$, $M=1.8$ on $16^3\times32$ lattice.}
\label{tab:anoqmass-pl}
\end{center}
\end{table}

\begin{table}[p]
  \leavevmode
\begin{center}
\begin{tabular}{l|lllll}
  $m_f$ & \multicolumn{5}{c}{$N_s$} \\
  & 10 & 20 & 30 & 40 & 50\\
\hline
& \multicolumn{5}{c}{$\beta=2.2$, $M=1.3$ on $12^3\times24$}\\
\hline
$0.00$ & $-$ & $-$ & $-$ & $0.01565$ ($90$) & $0.01260$ ($81$)\cr
$0.03$ & $-$ & $-$ & $-$ & $0.01605$ ($84$) & $0.01324$ ($78$)\cr
$0.05$ & $-$ & $-$ & $-$ & $0.01657$ ($86$) & $0.01385$ ($80$)\cr
$0.1$ & $-$ & $-$ & $-$ & $0.0172$ ($11$) & $0.01491$ ($84$)\cr
\hline
& \multicolumn{5}{c}{$\beta=2.2$, $M=1.7$ on $12^3\times24$}\\
\hline
$0.00$ & $-$ & $-$ & $-$ & $0.00413$ ($23$) & $0.00319$ ($26$)\cr
$0.03$ & $-$ & $-$ & $-$ & $0.00399$ ($20$) & $0.00302$ ($24$)\cr
$0.05$ & $-$ & $-$ & $-$ & $0.00403$ ($25$) & $0.00286$ ($25$)\cr
$0.1$ & $-$ & $-$ & $-$ & $0.00377$ ($21$) & $0.00257$ ($26$)\cr
\hline
& \multicolumn{5}{c}{$\beta=2.2$, $M=2.1$ on $12^3\times24$}\\
\hline
$0.00$ & $-$ & $-$ & $-$ & $0.00378$ ($25$) & $0.00303$ ($24$)\cr
$0.03$ & $-$ & $-$ & $-$ & $0.00386$ ($21$) & $0.00306$ ($19$)\cr
$0.05$ & $-$ & $-$ & $-$ & $0.00373$ ($22$) & $0.00299$ ($17$)\cr
$0.1$ & $-$ & $-$ & $-$ & $0.00388$ ($23$) & $0.00304$ ($24$)\cr
\end{tabular}
\caption{Anomalous quark mass for the RG improved action in the
 strong coupling region at $\beta=2.2$, $M=1.3,1.7,2.1$ on $12^3\times24$ 
lattice.}
\label{tbl:anoq-rg-st12M13}
\end{center}
\end{table}

\begin{table}[p]
  \leavevmode
\begin{center}
\begin{tabular}{l|lllll}
  $m_f$ & \multicolumn{5}{c}{$N_s$} \\
   & 10 & 20 & 30 & 40 & 50\\
\hline 
& \multicolumn{5}{c}{$\beta=2.2$, $M=1.3$ on $16^3\times24$}\\
\hline
$0.00$ & $0.09729$ ($99$) & $-$ & $0.02340$ ($41$) & $-$ & $0.01329$ ($48$)\cr
$0.03$ & $0.095$ ($1$) & $-$ & $0.02345$ ($39$) & $-$ & $0.01356$ ($49$)\cr
$0.05$ & $0.0928$ ($11$) & $-$ & $0.02340$ ($41$) & $-$ & $0.01378$ ($59$)\cr
$0.1$ & $0.0890$ ($11$) & $-$ & $0.02350$ ($55$) & $-$ & $0.01422$ ($75$)\cr
\hline 
& \multicolumn{5}{c}{$\beta=2.2$, $M=1.7$ on $16^3\times24$}\\
\hline
$0.00$ & $0.02486$ ($42$) & $0.00966$ ($35$) & $0.00626$ ($39$)
 & $0.00443$ ($28$) & $0.00344$ ($33$)\cr
$0.03$ & $0.02387$ ($38$) & $0.00920$ ($32$) & $0.00586$ ($35$)
 & $0.00412$ ($26$) & $0.00329$ ($28$)\cr
$0.05$ & $0.02281$ ($35$) & $0.00875$ ($33$) & $0.00549$ ($33$)
 & $0.00379$ ($23$) & $0.00321$ ($22$)\cr
$0.1$ & $0.02113$ ($35$) & $0.00800$ ($32$) & $0.00480$ ($33$)
 & $0.00326$ ($24$) & $0.00296$ ($20$)\cr
\hline 
& \multicolumn{5}{c}{$\beta=2.2$, $M=2.1$ on $16^3\times24$}\\
\hline
$0.00$ & $0.01633$ ($35$) & $0.00788$ ($30$) & $0.00490$ ($26$)
 & $0.00377$ ($25$) & $0.00314$ ($22$)\cr
$0.03$ & $0.01626$ ($29$) & $0.00804$ ($28$) & $0.00511$ ($21$)
 & $0.00381$ ($19$) & $0.00318$ ($18$)\cr
$0.05$ & $0.0158$ ($3$) & $0.00782$ ($29$) & $0.00501$ ($20$)
 & $0.00368$ ($18$) & $0.00312$ ($16$)\cr
$0.1$ & $0.01570$ ($28$) & $0.00811$ ($30$) & $0.00535$ ($21$)
 & $0.00374$ ($20$) & $0.00317$ ($19$)\cr
\end{tabular}
\caption{Anomalous quark mass for the RG improved action in the
 strong coupling region at $\beta=2.2$, $M=1.3,1.7,2.1$ on $16^3\times24$ 
lattice.}
\label{tbl:anoq-rg-st16M13}
\end{center}
\end{table}

\begin{table}[p]
  \leavevmode
\begin{center}
\begin{tabular}{l|lllll}
$m_f$  & \multicolumn{5}{c}{$N_s$} \\
   & 4 & 10 & 16 & 20 & 24\\
\hline 
$0.0$ & $0.01649$ ($11$) & $0.000817$ ($20$) & $0.000146$ ($22$)
 & $0.000042$ ($11$) & $0.000009$ ($2$)\cr
$0.02$ & $0.016238$ ($99$) & $0.000812$ ($17$) & $0.000136$ ($17$)
 & $0.0000405$ ($87$) & $0.0000087$ ($15$)\cr
$0.04$ & $0.015920$ ($80$) & $0.000803$ ($16$) & $0.000116$ ($12$)
 & $0.0000397$ ($78$) & $0.0000090$ ($19$)\cr
$0.06$ & $0.015677$ ($70$) & $0.000801$ ($16$) & $0.0001060$ ($76$)
 & $0.0000387$ ($73$) & $0.0000089$ ($19$)\cr
\end{tabular}
\caption{Anomalous quark mass for the RG improved action in the
 weak coupling region at
 $\beta=2.6$, $M=1.8$ on $16^3\times32$ lattice.}
\label{tab:anoqmass-rg}
\end{center}
\end{table}

\begin{table}[p]
  \leavevmode
\begin{center}
\begin{tabular}{l|lllll}
$m_f$ & \multicolumn{5}{c}{$N_s$} \\
      & 10 & 20 & 30 & 40 & 50\\
\hline 
& \multicolumn{5}{c}{$\beta=5.65$, $M=1.3$ on $12^3\times24$}\\
\hline
$0.00$ & $0.4982$ ($80$) & $0.2504$ ($83$) & $0.1618$ ($47$)
 & $0.1234$ ($71$) & $0.1007$ ($59$)\cr
$0.03$ & $0.5817$ ($81$) & $0.3197$ ($82$) & $0.2258$ ($45$)
 & $0.1845$ ($66$) & $0.1597$ ($61$)\cr
$0.04$ & $0.6078$ ($81$) & $0.3415$ ($82$) & $0.2462$ ($44$)
 & $0.2038$ ($63$) & $0.1787$ ($64$)\cr
$0.05$ & $0.6341$ ($82$) & $0.3635$ ($83$) & $0.2666$ ($43$)
 & $0.2233$ ($61$) & $0.1978$ ($66$)\cr
$0.065$ & $0.6760$ ($83$) & $0.3981$ ($84$) & $0.2987$ ($44$)
 & $0.2536$ ($61$) & $0.2272$ ($71$)\cr
$0.075$ & $0.7030$ ($83$) & $0.4206$ ($84$) & $0.3195$ ($43$)
 & $0.2736$ ($59$) & $0.2466$ ($72$)\cr
$0.1$ & $0.7733$ ($84$) & $0.4790$ ($85$) & $0.3733$ ($43$)
 & $0.3249$ ($59$) & $0.2964$ ($76$)\cr
\hline 
& \multicolumn{5}{c}{$\beta=5.65$, $M=1.7$ on $12^3\times24$}\\
\hline
$0.00$ & $0.2675$ ($55$) & $0.1292$ ($68$) & $0.0889$ ($61$)
 & $0.0833$ ($55$) & $0.0537$ ($50$)\cr
$0.03$ & $0.4190$ ($55$) & $0.2840$ ($61$) & $0.2440$ ($53$)
 & $0.2384$ ($54$) & $0.2114$ ($50$)\cr
$0.04$ & $0.4667$ ($57$) & $0.3329$ ($61$) & $0.2927$ ($53$)
 & $0.2882$ ($54$) & $0.2630$ ($54$)\cr
$0.05$ & $0.5148$ ($59$) & $0.3822$ ($61$) & $0.3416$ ($52$)
 & $0.3375$ ($52$) & $0.3147$ ($58$)\cr
$0.065$ & $0.5912$ ($62$) & $0.4589$ ($63$) & $0.4193$ ($55$)
 & $0.4171$ ($61$) & $0.3929$ ($64$)\cr
$0.075$ & $0.6401$ ($63$) & $0.5094$ ($62$) & $0.4692$ ($54$)
 & $0.4662$ ($59$) & $0.4453$ ($65$)\cr
$0.1$ & $0.7679$ ($65$) & $0.6401$ ($61$) & $0.6004$ ($56$)
 & $0.5966$ ($61$) & $0.5781$ ($69$)\cr
\hline 
& \multicolumn{5}{c}{$\beta=5.65$, $M=2.1$ on $12^3\times24$}\\
\hline
$0.00$ & $0.2418$ ($68$) & $0.1393$ ($61$) & $0.1067$ ($60$)
 & $0.0876$ ($68$) & $0.0839$ ($67$)\cr
$0.03$ & $0.4143$ ($70$) & $0.3206$ ($56$) & $0.2903$ ($55$)
 & $0.2719$ ($64$) & $0.2689$ ($64$)\cr
$0.04$ & $0.4722$ ($74$) & $0.3813$ ($57$) & $0.3521$ ($55$)
 & $0.3344$ ($65$) & $0.3327$ ($67$)\cr
$0.05$ & $0.5295$ ($77$) & $0.4416$ ($57$) & $0.4134$ ($55$)
 & $0.3962$ ($64$) & $0.3955$ ($69$)\cr
$0.065$ & $0.6175$ ($82$) & $0.5328$ ($62$) & $0.5056$ ($60$)
 & $0.4895$ ($69$) & $0.4888$ ($77$)\cr
$0.075$ & $0.6740$ ($83$) & $0.5929$ ($60$) & $0.5666$ ($58$)
 & $0.5505$ ($67$) & $0.5505$ ($76$)\cr
$0.1$ & $0.8172$ ($85$) & $0.7439$ ($58$) & $0.7195$ ($57$)
 & $0.7037$ ($68$) & $0.7037$ ($77$)\cr
\hline 
& \multicolumn{5}{c}{$\beta=5.65$, $M=2.5$ on $12^3\times24$}\\
\hline
$0.00$ & $0.2624$ ($63$) & $0.1701$ ($48$) & $0.1546$ ($43$) & $-$
 & $0.1430$ ($83$)\cr
$0.03$ & $0.4007$ ($59$) & $0.3049$ ($49$) & $0.2796$ ($43$) & $-$
 & $0.2439$ ($77$)\cr
$0.04$ & $0.4496$ ($59$) & $0.3535$ ($52$) & $0.3270$ ($46$) & $-$
 & $0.2894$ ($80$)\cr
$0.05$ & $0.4976$ ($59$) & $0.4012$ ($56$) & $0.3729$ ($49$) & $-$
 & $0.3320$ ($83$)\cr
$0.065$ & $0.5682$ ($61$) & $0.4698$ ($61$) & $0.4385$ ($54$) & $-$
 & $0.3880$ ($88$)\cr
$0.075$ & $0.6146$ ($60$) & $0.5153$ ($63$) & $0.4806$ ($56$) & $-$
 & $0.4236$ ($89$)\cr
$0.1$ & $0.7284$ ($60$) & $0.6239$ ($68$) & $0.5783$ ($60$) & $-$
 & $0.4984$ ($88$)\cr
\end{tabular}
\caption{Pion mass squared for the plaquette action in the strong coupling
 region at $\beta=5.65$, $M=1.3,1.7,2.1,2.5$ on $12^3\times24$ lattice.}
\label{tbl:pion-pl-st12M13}
\end{center}
\end{table}

\begin{table}[p]
  \leavevmode
\begin{center}
\begin{tabular}{l|lllll}
$m_f$ & \multicolumn{5}{c}{$N_s$} \\
      & 10 & 20 & 30 & 40 & 50\\
\hline 
& \multicolumn{5}{c}{$\beta=5.65$, $M=1.3$ on $16^3\times24$}\\
\hline
$0.00$ & $0.499$ ($5$) & $0.2556$ ($40$) & $-$ & $-$ & $0.0957$ ($53$)\cr
$0.03$ & $0.5823$ ($52$) & $0.3257$ ($42$) & $-$ & $-$ & $0.1545$ ($50$)\cr
$0.04$ & $0.6084$ ($53$) & $0.3480$ ($43$) & $-$ & $-$ & $0.1734$ ($49$)\cr
$0.05$ & $0.6347$ ($54$) & $0.3704$ ($45$) & $-$ & $-$ & $0.1925$ ($48$)\cr
$0.065$ & $0.6766$ ($55$) & $0.4056$ ($47$) & $-$ & $-$ & $0.2215$ ($49$)\cr
$0.075$ & $0.7035$ ($56$) & $0.4284$ ($48$) & $-$ & $-$ & $0.2409$ ($48$)\cr
$0.1$ & $0.7737$ ($58$) & $0.4876$ ($50$) & $-$ & $-$ & $0.2905$ ($46$)\cr
\hline 
& \multicolumn{5}{c}{$\beta=5.65$, $M=1.7$ on $16^3\times24$}\\
\hline
$0.00$ & $0.2661$ ($49$) & $0.1358$ ($56$) & $0.0888$ ($36$)
 & $0.0599$ ($38$) & $0.0520$ ($31$)\cr
$0.03$ & $0.4179$ ($51$) & $0.2916$ ($55$) & $0.2434$ ($34$)
 & $0.2183$ ($32$) & $0.2086$ ($30$)\cr
$0.04$ & $0.4656$ ($53$) & $0.3411$ ($57$) & $0.2929$ ($37$)
 & $0.2693$ ($33$) & $0.2599$ ($35$)\cr
$0.05$ & $0.5137$ ($55$) & $0.3911$ ($59$) & $0.3426$ ($40$)
 & $0.3204$ ($33$) & $0.3113$ ($38$)\cr
$0.065$ & $0.5898$ ($58$) & $0.4687$ ($65$) & $0.4198$ ($46$)
 & $0.3992$ ($37$) & $0.3894$ ($47$)\cr
$0.075$ & $0.6392$ ($59$) & $0.5198$ ($64$) & $0.4707$ ($48$)
 & $0.4512$ ($36$) & $0.4414$ ($48$)\cr
$0.1$ & $0.7676$ ($61$) & $0.6513$ ($66$) & $0.6019$ ($53$)
 & $0.5849$ ($36$) & $0.5735$ ($54$)\cr
\hline 
& \multicolumn{5}{c}{$\beta=5.65$, $M=2.1$ on $16^3\times24$}\\
\hline
$0.00$ & $0.2317$ ($40$) & $0.1434$ ($33$) & $0.1132$ ($72$)
 & $0.0935$ ($50$) & $0.0815$ ($38$)\cr
$0.03$ & $0.4052$ ($46$) & $0.3247$ ($37$) & $0.2973$ ($62$)
 & $0.2792$ ($45$) & $0.2675$ ($43$)\cr
$0.04$ & $0.4629$ ($50$) & $0.3860$ ($42$) & $0.3604$ ($63$)
 & $0.3416$ ($49$) & $0.3317$ ($47$)\cr
$0.05$ & $0.5202$ ($55$) & $0.4464$ ($46$) & $0.4226$ ($62$)
 & $0.4034$ ($51$) & $0.3945$ ($51$)\cr
$0.065$ & $0.6085$ ($62$) & $0.5390$ ($52$) & $0.5166$ ($68$)
 & $0.4979$ ($58$) & $0.4902$ ($60$)\cr
$0.075$ & $0.6652$ ($65$) & $0.5982$ ($56$) & $0.5774$ ($63$)
 & $0.5589$ ($56$) & $0.5509$ ($61$)\cr
$0.1$ & $0.8095$ ($72$) & $0.7484$ ($60$) & $0.7299$ ($60$)
 & $0.7131$ ($54$) & $0.7036$ ($66$)\cr
\end{tabular}
\caption{Pion mass squared for the plaquette action in the strong coupling
 region at $\beta=5.65$, $M=1.3,1.7,2.1$ on $16^3\times24$ lattice.}
\label{tbl:pion-pl-st16M13}
\end{center}
\end{table}

\begin{table}[p]
  \leavevmode
\begin{center}
\begin{tabular}{l|lllll}
$m_f$ & \multicolumn{5}{c}{$N_s$} \\
      & 10 & 20 & 30 & 40 & 50\\
\hline 
& \multicolumn{5}{c}{$\beta=2.2$, $M=1.3$ on $12^3\times24$}\\
\hline
$0.00$ & $0.3038$ ($55$) & $0.1208$ ($55$) & $0.0684$ ($71$)
 & $0.0415$ ($50$) & $0.0389$ ($60$)\cr
$0.03$ & $0.4002$ ($52$) & $0.2106$ ($52$) & $0.1579$ ($64$)
 & $0.1325$ ($47$) & $0.1253$ ($57$)\cr
$0.04$ & $0.4301$ ($52$) & $0.2388$ ($53$) & $0.1867$ ($63$)
 & $0.1619$ ($49$) & $0.1518$ ($59$)\cr
$0.05$ & $0.4604$ ($52$) & $0.2671$ ($54$) & $0.2154$ ($61$)
 & $0.1916$ ($49$) & $0.1786$ ($60$)\cr
$0.065$ & $0.5083$ ($52$) & $0.3120$ ($56$) & $0.2605$ ($64$)
 & $0.2363$ ($55$) & $0.2215$ ($64$)\cr
$0.075$ & $0.5395$ ($52$) & $0.3410$ ($56$) & $0.2894$ ($61$)
 & $0.2667$ ($53$) & $0.2494$ ($63$)\cr
$0.1$ & $0.6209$ ($51$) & $0.4169$ ($56$) & $0.3647$ ($58$)
 & $0.3433$ ($54$) & $0.3228$ ($60$)\cr
\hline 
& \multicolumn{5}{c}{$\beta=2.2$, $M=1.7$ on $12^3\times24$}\\
\hline
$0.00$ & $0.1304$ ($50$) & $0.0575$ ($50$) & $0.0405$ ($57$)
 & $0.0408$ ($42$) & $0.0302$ ($73$)\cr
$0.03$ & $0.2892$ ($51$) & $0.2186$ ($46$) & $0.2013$ ($54$)
 & $0.2049$ ($39$) & $0.1898$ ($57$)\cr
$0.04$ & $0.3394$ ($54$) & $0.2697$ ($48$) & $0.2529$ ($56$)
 & $0.2588$ ($40$) & $0.2410$ ($55$)\cr
$0.05$ & $0.3900$ ($57$) & $0.3210$ ($49$) & $0.3042$ ($58$)
 & $0.3119$ ($40$) & $0.2917$ ($51$)\cr
$0.065$ & $0.4695$ ($64$) & $0.4016$ ($55$) & $0.3850$ ($66$)
 & $0.3961$ ($48$) & $0.3724$ ($56$)\cr
$0.075$ & $0.5213$ ($64$) & $0.4540$ ($54$) & $0.4372$ ($68$)
 & $0.4486$ ($46$) & $0.4234$ ($53$)\cr
$0.1$ & $0.6559$ ($67$) & $0.5904$ ($56$) & $0.5737$ ($77$)
 & $0.5863$ ($47$) & $0.5582$ ($58$)\cr
\hline 
& \multicolumn{5}{c}{$\beta=2.2$, $M=2.1$ on $12^3\times24$}\\
\hline
$0.00$ & $0.1159$ ($64$) & $0.0810$ ($63$) & $0.0645$ ($68$)
 & $0.0556$ ($73$) & $0.0415$ ($82$)\cr
$0.03$ & $0.2879$ ($60$) & $0.2510$ ($55$) & $0.2303$ ($64$)
 & $0.2272$ ($71$) & $0.2131$ ($73$)\cr
$0.04$ & $0.3461$ ($63$) & $0.3083$ ($56$) & $0.2873$ ($66$)
 & $0.2848$ ($75$) & $0.2705$ ($74$)\cr
$0.05$ & $0.4038$ ($65$) & $0.3644$ ($58$) & $0.3429$ ($68$)
 & $0.3415$ ($78$) & $0.3274$ ($72$)\cr
$0.065$ & $0.4911$ ($72$) & $0.4519$ ($64$) & $0.4276$ ($76$)
 & $0.4292$ ($87$) & $0.4143$ ($81$)\cr
$0.075$ & $0.548$ ($7$) & $0.5068$ ($66$) & $0.4817$ ($73$)
 & $0.4851$ ($85$) & $0.4706$ ($76$)\cr
$0.1$ & $0.6906$ ($72$) & $0.6482$ ($71$) & $0.6198$ ($69$)
 & $0.6279$ ($84$) & $0.6136$ ($76$)\cr
\hline 
& \multicolumn{5}{c}{$\beta=2.2$, $M=2.5$ on $12^3\times24$}\\
\hline
$0.00$ & $0.1549$ ($48$) & $0.102$ ($6$) & $0.1127$ ($63$) & $-$ & $-$\cr
$0.03$ & $0.2814$ ($47$) & $0.2208$ ($57$) & $0.2184$ ($53$) & $-$ & $-$\cr
$0.04$ & $0.3256$ ($48$) & $0.2649$ ($57$) & $0.2611$ ($54$) & $-$ & $-$\cr
$0.05$ & $0.3692$ ($51$) & $0.3076$ ($58$) & $0.3018$ ($53$) & $-$ & $-$\cr
$0.065$ & $0.4329$ ($56$) & $0.3699$ ($59$) & $0.3607$ ($54$) & $-$ & $-$\cr
$0.075$ & $0.4754$ ($59$) & $0.4099$ ($59$) & $0.3965$ ($52$) & $-$ & $-$\cr
$0.1$ & $0.5793$ ($65$) & $0.5058$ ($58$) & $0.4797$ ($48$) & $-$ & $-$\cr
\end{tabular}
\caption{Pion mass squared for the RG improved action in the strong
 coupling region at $\beta=2.2$, $M=1.3,1.7,2.1,2.5$ on $12^3\times24$ 
lattice.}
\label{tbl:pion-rg-st12M13}
\end{center}
\end{table}

\begin{table}[p]
  \leavevmode
\begin{center}
\begin{tabular}{l|lllll}
$m_f$ & \multicolumn{5}{c}{$N_s$} \\
      & 10 & 20 & 30 & 40 & 50\\
\hline 
& \multicolumn{5}{c}{$\beta=2.2$, $M=1.3$ on $16^3\times24$}\\
\hline
$0.00$ & $0.3126$ ($54$) & $-$ & $0.0709$ ($19$) & $-$ & $0.0464$ ($30$)\cr
$0.03$ & $0.4095$ ($56$) & $-$ & $0.1609$ ($19$) & $-$ & $0.1348$ ($28$)\cr
$0.04$ & $0.4400$ ($58$) & $-$ & $0.1900$ ($19$) & $-$ & $0.1634$ ($29$)\cr
$0.05$ & $0.4707$ ($59$) & $-$ & $0.2190$ ($21$) & $-$ & $0.1918$ ($30$)\cr
$0.065$ & $0.5194$ ($62$) & $-$ & $0.2642$ ($25$) & $-$ & $0.2368$ ($34$)\cr
$0.075$ & $0.5508$ ($63$) & $-$ & $0.2938$ ($26$) & $-$ & $0.2652$ ($34$)\cr
$0.1$ & $0.6326$ ($66$) & $-$ & $0.3702$ ($33$) & $-$ & $0.3395$ ($35$)\cr
\hline 
& \multicolumn{5}{c}{$\beta=2.2$, $M=1.7$ on $16^3\times24$}\\
\hline
$0.00$ & $0.1276$ ($49$) & $0.0579$ ($37$) & $0.0392$ ($36$)
 & $0.0308$ ($35$) & $0.0246$ ($35$)\cr
$0.03$ & $0.2882$ ($46$) & $0.2222$ ($34$) & $0.2012$ ($32$)
 & $0.1959$ ($33$) & $0.1908$ ($34$)\cr
$0.04$ & $0.3392$ ($45$) & $0.2763$ ($36$) & $0.2527$ ($35$)
 & $0.2498$ ($37$) & $0.2454$ ($36$)\cr
$0.05$ & $0.3905$ ($44$) & $0.3300$ ($38$) & $0.3042$ ($37$)
 & $0.3034$ ($40$) & $0.2995$ ($38$)\cr
$0.065$ & $0.4704$ ($46$) & $0.4131$ ($44$) & $0.3854$ ($42$)
 & $0.3874$ ($46$) & $0.3837$ ($44$)\cr
$0.075$ & $0.5228$ ($45$) & $0.4667$ ($45$) & $0.4381$ ($42$)
 & $0.4409$ ($48$) & $0.4378$ ($45$)\cr
$0.1$ & $0.6582$ ($45$) & $0.6046$ ($48$) & $0.5760$ ($47$)
 & $0.5798$ ($53$) & $0.5773$ ($47$)\cr
\hline 
& \multicolumn{5}{c}{$\beta=2.2$, $M=2.1$ on $16^3\times24$}\\
\hline
$0.00$ & $0.1194$ ($51$) & $0.0735$ ($40$) & $0.0539$ ($46$)
 & $0.0455$ ($41$) & $0.0492$ ($46$)\cr
$0.03$ & $0.2868$ ($42$) & $0.2442$ ($38$) & $0.2303$ ($45$)
 & $0.2177$ ($40$) & $0.2238$ ($44$)\cr
$0.04$ & $0.3425$ ($40$) & $0.302$ ($4$) & $0.2912$ ($49$)
 & $0.2754$ ($43$) & $0.2840$ ($48$)\cr
$0.05$ & $0.398$ ($4$) & $0.3590$ ($41$) & $0.3509$ ($51$)
 & $0.3324$ ($45$) & $0.3427$ ($49$)\cr
$0.065$ & $0.4826$ ($43$) & $0.4459$ ($47$) & $0.4413$ ($59$)
 & $0.4202$ ($53$) & $0.4336$ ($58$)\cr
$0.075$ & $0.5376$ ($43$) & $0.5019$ ($48$) & $0.4991$ ($58$)
 & $0.4763$ ($51$) & $0.4899$ ($56$)\cr
$0.1$ & $0.6770$ ($49$) & $0.6434$ ($54$) & $0.6445$ ($59$)
 & $0.6197$ ($51$) & $0.6336$ ($56$)\cr
\end{tabular}
\caption{Pion mass squared for the RG improved action in the strong
 coupling region at $\beta=2.2$, $M=1.3, 1.7, 2.1$ on $16^3\times24$
 lattice.} 
\label{tbl:pion-rg-st16M13}
\end{center}
\end{table}

\begin{table}[p]
  \leavevmode
\begin{center}
\begin{tabular}{l|lllll}
$m_f$ & \multicolumn{5}{c}{$N_s$} \\
      & 4 & 10 & 16 & 20 & 30\\
\hline 
$0.0$ & $0.0891$ ($49$) & $0.0096$ ($36$) & $0.0025$ ($27$)
 & $0.0063$ ($21$) & $0.0019$ ($20$)\cr
$0.02$ & $0.1617$ ($43$) & $0.0795$ ($30$) & $0.0711$ ($24$)
 & $0.0736$ ($19$) & $0.0703$ ($17$)\cr
$0.03$ & $0.1959$ ($41$) & $0.1130$ ($34$) & $0.1032$ ($25$)
 & $0.1051$ ($20$) & $0.1029$ ($21$)\cr
$0.04$ & $0.2310$ ($39$) & $0.1471$ ($32$) & $0.1359$ ($24$)
 & $0.1371$ ($20$) & $0.1362$ ($20$)\cr
$0.04$ & $0.2311$ ($40$) & $0.1471$ ($36$) & $0.1362$ ($26$)
 & $0.1380$ ($21$) & $0.1366$ ($23$)\cr
$0.05$ & $0.2671$ ($38$) & $0.1823$ ($35$) & $0.1710$ ($27$)
 & $0.1714$ ($23$) & $0.1706$ ($20$)\cr
$0.06$ & $0.3041$ ($36$) & $0.2179$ ($34$) & $0.2059$ ($27$)
 & $0.2061$ ($23$) & $0.2059$ ($20$)\cr
\end{tabular}
\caption{Pion mass squared for the plaquette action in the weak coupling
 region at
 $\beta=6.0$, $M=1.8$ on $16^3\times32$ lattice.}
\label{tab:pionmass-pl}
\end{center}
\end{table}

\begin{table}[p]
  \leavevmode
\begin{center}
\begin{tabular}{l|lllll}
$m_f$ & \multicolumn{5}{c}{$N_s$} \\
      & 4 & 10 & 16 & 20 & 24\\
\hline 
$0.0$ & $0.0536$ ($26$) & $0.0044$ ($33$) & $0.0052$ ($23$)
 & $0.0023$ ($19$) & $0.0022$ ($16$)\cr
$0.02$ & $0.1188$ ($24$) & $0.0670$ ($30$) & $0.0671$ ($23$)
 & $0.0646$ ($15$) & $0.0652$ ($14$)\cr
$0.03$ & $0.1505$ ($26$) & $0.0967$ ($32$) & $0.0969$ ($26$)
 & $0.0942$ ($17$) & $0.0959$ ($15$)\cr
$0.04$ & $0.1826$ ($29$) & $0.1268$ ($30$) & $0.1270$ ($27$)
 & $0.1243$ ($17$) & $0.1268$ ($15$)\cr
$0.04$ & $0.1835$ ($29$) & $0.1274$ ($34$) & $0.1280$ ($29$)
 & $0.1246$ ($19$) & $0.1274$ ($16$)\cr
$0.05$ & $0.2150$ ($36$) & $0.1583$ ($31$) & $0.1582$ ($31$)
 & $0.1559$ ($18$) & $0.1580$ ($16$)\cr
$0.06$ & $0.2491$ ($39$) & $0.1907$ ($33$) & $0.1904$ ($31$)
 & $0.1877$ ($18$) & $0.1908$ ($16$)\cr
\end{tabular}
\caption{Pion mass squared for the RG improved action in the weak
 coupling region at
 $\beta=2.6$, $M=1.8$ on $16^3\times32$ lattice.}
\label{tab:pionmass-rg}
\end{center}
\end{table}



\end{document}